\documentclass[a4paper,superscriptaddress,floatfix]{revtex4}

\usepackage{epsfig}
\usepackage{amsfonts}
\usepackage{amsmath}
\usepackage{amssymb}
\usepackage{amsthm}
\usepackage{float}
\usepackage{array}
\usepackage{booktabs}
\usepackage{color}
\usepackage{graphicx}
\usepackage{hyperref}
\usepackage{multirow}
\usepackage{slashed}
\usepackage{subfigure}
\usepackage{theorem}
\usepackage{textcomp}
\usepackage[utf8]{inputenc}

\allowdisplaybreaks[1]

\numberwithin{equation}{section}

\begin{document}

\begin{flushright}
 \hspace{3cm} MS-TP-18-15
\end{flushright}

\title{Neutral Current 
Forward-Backward Asymmetry:       
from $\theta_W$ to PDF Determinations}

\author{E. Accomando}
\email[E-mail: ]{e.accomando@soton.ac.uk}
\affiliation{School of Physics \& Astronomy, University of Southampton, Highfield, Southampton SO17 1BJ, UK}
\affiliation{Particle Physics Department, Rutherford Appleton Laboratory, Chilton, Didcot, Oxon OX11 0QX, UK}

\author{J. Fiaschi}
\email[E-mail: ]{fiaschi@uni-muenster.de}
\affiliation{School of Physics \& Astronomy, University of Southampton, Highfield, Southampton SO17 1BJ, UK}
\affiliation{Particle Physics Department, Rutherford Appleton Laboratory, Chilton, Didcot, Oxon OX11 0QX, UK}
\affiliation{Institut f{\" u}r Theoretische Physik, Universit{\" a}t M{\" u}nster, D 48149 M{\" u}nster, Germany}

\author{F. Hautmann}
\email[E-mail: ]{hautmann@thphys.ox.ac.uk}
\affiliation{Particle Physics Department, Rutherford Appleton Laboratory, Chilton, Didcot, Oxon OX11 0QX, UK}
\affiliation{Elementaire Deeltjes Fysica, Universiteit Antwerpen, B 2020 Antwerpen, Belgium}
\affiliation{Theoretical Physics Department, University of Oxford, Oxford OX1 3NP, UK}

\author{S. Moretti}
\email[E-mail: ]{s.moretti@soton.ac.uk}
\affiliation{School of Physics \& Astronomy, University of Southampton, Highfield, Southampton SO17 1BJ, UK}
\affiliation{Particle Physics Department, Rutherford Appleton Laboratory, Chilton, Didcot, Oxon OX11 0QX, UK}

 \begin{abstract}
{
Measurements of the forward-backward asymmetry in neutral-current Drell-Yan di-lepton production have primarily been used for determinations of the weak mixing angle $\theta_W$.
We observe that, unlike the case of Run-I of the Large Hadron Collider (LHC Run-I), for the first time at the LHC Run-II the reconstructed forward-backward asymmetry
has the capability of placing useful constraints on the determination of the parton distribution functions (PDFs).
By examining the statistical and the PDF uncertainties on the reconstructed forward-backward asymmetry, we investigate its potential for disentangling the flavour content of quark and antiquark PDFs.
Access to the valence/sea $u$-quark and to the sea up-type antiquark PDFs, in particular, may be gained by the appropriate use of selection cuts in the rapidity of the emerging lepton pair
in regions both near the $Z$-boson peak and away from it, in a manner complementary, though more indirect, to the case of the charged-current asymmetry.
We study the extension of these results for the planned high-luminosity (HL) LHC.
}

\end{abstract}

\maketitle

\setcounter{footnote}{0}

\section{Introduction}

The forward-backward asymmetry in neutral-current (NC) Drell-Yan (DY) lepton pair production has traditionally been used at hadron-hadron colliders in the context of precision electroweak measurements,
in particular for determinations of the weak mixing angle $\theta_W$~\cite{Aad:2015uau,Chatrchyan:2011ya,Aaij:2015lka}.
However, mostly due to the theoretical systematics coming from the parton distribution functions (PDFs), the accuracy reached in $\theta_W$ determinations from measurements of the forward-backward asymmetry
at the Large Hadron Collider (LHC) is currently an order of magnitude lower than in the case of $\theta_W$ determinations from lepton colliders~\cite{ALEPH:2005ab}. 

In Ref.~\cite{Accomando:2017scx} we have proposed that measurements of the NC forward-backward asymmetry (henceforth denoted as $A_{\rm FB}$) at the LHC Run-II (as well as in future high-luminosity runs at the HL-LHC) 
can usefully be employed in PDF determinations, in a fashion complementary to measurements of the charged-current (CC) asymmetry, which have long been used in PDF fits to access the $d/u$ quark PDF ratio
(see~\cite{Dulat:2015mca,Ball:2017nwa,Alekhin:2017kpj,Harland-Lang:2014zoa,Abramowicz:2015mha,Aaboud:2016btc} and references therein).
While CC and NC doubly differential cross sections in the DY di-lepton mass and rapidity have extensively been used for PDF determinations, Ref.~\cite{Accomando:2017scx} investigates the role of the angular information encoded in the 
$A_{\rm FB}$, which is related to the single-lepton pseudorapidity and, once combined with di-lepton mass and rapidity, would qualitatively correspond to triply differential cross sections.
Precision measurements of such cross sections~\cite{Aaboud:2017ffb} have not yet been exploited systematically for PDF determinations.
It is noted in~\cite{Accomando:2017scx} that, once mapped in the invariant mass of the di-lepton final state, both the NC differential cross section and the $A_{\rm FB}$ display statistical errors which, with the luminosity reachable 
at Run-II, become competitive with those assigned to quark PDFs in a broad kinematic range around the $Z$-boson peak. 

In this paper we follow the proposal~\cite{Accomando:2017scx} and explore the NC asymmetry focusing on its potential for disentangling the flavour content of PDFs.
Specifically, we find that it is possible to probe the quark PDF flavour structure with the $A_{\rm FB}$ through a judicious use of selection cuts in the rapidity of the emerging di-lepton pair,
whose four-momentum is completely accessible in NC DY, unlike the case of the final state of CC DY which contains a neutrino.
Although access to the $d/u$ quark PDF ratio is more indirect with the NC $A_{\rm FB}$ than with the CC asymmetry, the PDF kinematical region in longitudinal momentum fraction $x$ and evolution scale $Q^2$ is larger in the NC case,
as both on-peak and off-peak cross sections can be used, whilst in the CC case the escaping neutrino prevents the reconstruction of the $W^\pm$ mass so that the $W^\pm$ peak region necessarily dominates the cross section. 
We observe that selecting high di-lepton rapidities enhances the contribution of higher-order radiative QCD effects~\cite{Dooling:2012uw,Hautmann:2012dw} and subleading production channels~\cite{Hautmann:2012sh,Hautmann:2012pf}
so that this region also becomes sensitive to the gluon PDF.

The paper is organised as follows.
In Sec.~\ref{sec:AFB} we briefly summarise the set-up of the calculation of Ref.~\cite{Accomando:2017scx} (which was applied to the CT14nnlo~\cite{Dulat:2015mca} and NNPDF3.1~\cite{Ball:2017nwa} PDF sets)
and we extend its results to the case of the ABMP16nnlo~\cite{Alekhin:2017kpj}, HERA2.0nnlo~\cite{Abramowicz:2015mha} and MMHT2014nnlo~\cite{Harland-Lang:2014zoa} PDF sets. 
In Sec.~\ref{sec:parton_luminosities} we address the question of flavour content and investigate parton luminosities in the kinematic regions relevant for the $A_{\rm FB}$ around the $Z$-boson peak.
In Sec.~\ref{sec:sensitivity} we analyse the $A_{\rm FB}$ in the high-rapidity region and we study the role of rapidity selection cuts to disentangle the PDF flavour structure. 
We give conclusions in Sec.~\ref{sec:summa}.

\section{The Differential 
Forward-Backward Asymmetry}
\label{sec:AFB}

The forward-backward asymmetry $A_{\rm FB}$ in the $\gamma^*,Z$ mediated DY production process $pp\to \gamma^*,Z\to \ell^+\ell^-$ (with $\ell=e,\mu$) is given by 
\begin{equation}
\label{AFB-NC}
  A_{\rm FB} = \frac { d \sigma / d M(\ell^+\ell^-)[\eta(\ell^+)>0] - d \sigma / d M(\ell^+\ell^-)[\eta(\ell^+)<0] }      { d \sigma / d M(\ell^+\ell^-)[\eta(\ell^+)>0] + d \sigma / d M(\ell^+\ell^-)[\eta(\ell^+)<0] },
\end{equation}
where $M(\ell^+\ell^-)$ is the DY lepton pair invariant mass, $\eta(\ell^+)$ is charged lepton pseudorapidity while the identification of the Forward (F) and Backward (B) hemispheres via the restrictions
$\eta(\ell^+)>0$ and $\eta(\ell^+)<0$ is obtained by exploiting the di-lepton boost variable as discussed in~\cite{Dittmar:1996my,Rizzo:2009pu,Accomando:2015cfa,Accomando:2016tah,Accomando:2016ehi}.
Hence, this is effectively a `di-lepton FB asymmetry'. 

This asymmetry is to be contrasted with the `lepton charge asymmetry' in the CC channel, defined as 
\begin{equation}
\label{AFB-CC}
 A_{\rm CC} = \frac { d \sigma / d \eta(\ell^+) - d \sigma / d \eta(\ell^-) } { d \sigma / d \eta(\ell^+) + d \sigma / d \eta(\ell^-) }, 
\end{equation}
which has traditionally been used to constrain PDFs, particularly the ratio of $d$-quark to $u$-quark distributions.
From the charged current data, different observables can be actually reconstructed.
The two cross sections $\sigma (u\bar d\rightarrow \ell^+\nu$) and $\sigma (d\bar u\rightarrow \ell^-\bar  \nu$) give rise to different distributions owing to the different valence $u$ and $d$-quarks that dominate their production.
The distribution in the rapidity of the mediator, $W^\pm$, is governed by a sea quark and a sea anti-quark at low values.
Only a tiny difference is thus expected between the two above-mentioned cross sections. Instead, at large values of $y(W^\pm )$, the process is mainly initiated by a valence quark and a sea anti-quark.
A much larger difference is therefore expected between the two charged current cross sections.
This feature should enable one to distinguish between valence/sea u/d quark PDFs.
This highlights the potential role of a binning (even a cut) on the rapidity of the di-lepton system for charged current processes and in general.
Starting from the kinematic relations
\begin{equation}
M_W^2 = sx_1x_2 \  ,  \ \ \ \ \ \ \ \ y = {1\over 2}\ln \left ({x_1\over x_2}\right )
\end{equation}
where $\sqrt{s} = E_{coll}$ = 13 TeV and $x_i$ are the quark  
longitudinal-momentum fractions, the $W^\pm$ rapidity distributions can be directly related to the quark and anti-quark fractional momenta thus constraining the corresponding PDFs.
For $|y(W^\pm )|\simeq 2.5$, the measurable region at the $W$-boson peak 
would correspond to $x_1\simeq 0.1$ and 
$x_2\simeq 3\cdot 10^{-4}$ (low-$x$ anti-quark and large-$x$ quark 
as the charged lepton is likely to be emitted in the same direction as the valence quark).
This range could be used to constrain the PDFs and determine the parton luminosities. 

For the $CC$ leptonic channel, the $W^\pm$-rapidity can be derived using the reconstruction of the neutrino on the $W$-boson peak, $\sqrt{\hat s}\simeq M_W$.
This reconstruction algorithm however suffers from uncertainties. A better observable is the charged lepton pseudo-rapidity that is directly related to the scattering angle, $\eta(l^\pm ) = -\ln (\tan\theta /2)$.
The difference between the two processes, $q\bar q\rightarrow W^\pm\rightarrow l^\pm\nu_l$, is then enhanced at high values of $|\eta (l^\pm )|$, still  compatible with the detector acceptance. 

In order to reduce the systematics, including the theoretical uncertainty due to the higher order QCD and EW corrections, it is a standard practice to construct cross section ratios.
That is the reason to define the lepton charge asymmetry in Eq.~\ref{AFB-CC}.
This strategy is also recommended  to reduce the theoretical error due to the PDFs determination.
The PDFs uncertainty indeed cancels out, to a large extent, when considering cross section ratios.
This observation is the kernel of our analysis, which addresses the neutral current DY processes $q\bar q\rightarrow Z, \gamma^*\rightarrow l^+l^-$. 

This process is used to analyse the $Q^2$ dependence of the PDFs. While the CC process is sensitive to the sea anti-quark and the valence quark, the NC process encodes information 
about the sum of the sea $\bar d$, $\bar u$ anti-quarks and the valence $u$, $d$ quarks.
In contrast to the CC differential cross sections that appear in ratios, the NC ones are not normalised usually.
Recently,   LHC NC data have been released~\cite{Aaboud:2017ffb} in the form of triple differential cross sections in the di-lepton rapidity, di-lepton invariant mass and lepton scattering angle.
This powerful information however suffers from experimental and theoretical systematical errors.
In particular, the theoretical PDF’s uncertainty can be so large to overcome the potential differences coming from the comparison of the NC cross sections obtained using different PDF’s sets.
This would  weaken  the sensitivity of the LHC measurements to the $q$ and $\bar q$ PDFs.
At present, the discrepancy between the predictions from any two different PDF sets is compared to the statistical precision.
At the $W^\pm$ or $Z$-boson peaks, the cross sections are very high.
The signal is clear and the background is rather low.
That is why a sensitivity to small differences in the PDF’s parameterisation is claimed to be at hand. 

However, in our view, the PDF’s uncertainty could still represent a real obstacle.
This is the main reason for proposing the  use of the Forward-Backward asymmetry, which is a ratio of (differential) cross sections in the NC channel.  
It was observed in~\cite{Accomando:2017scx}, using the method of~\cite{Accomando:2015cfa,Accomando:2016tah} to define the reconstructed FB asymmetry, $A^*_{\rm FB}$, and 
to evaluate its PDF uncertainty, that whilst no gain in quark PDF determinations should be expected from the reconstruction of this observable with the data collected at 
$\sqrt{s}$ = 8 TeV, the scenario changes at the LHC Run-II and at the planned HL-LHC. 
For instance, for the projected luminosity $L =$ 3 ab$^{-1}$ of the HL LHC~\cite{Gianotti:2002xx}, and for the CT14~\cite{Dulat:2015mca} and NNPDF3.1~\cite{Ball:2017nwa} 
PDF sets, the statistical error is found~\cite{Accomando:2017scx} to be lower than the PDF error in the low to intermediate mass region $M_{\ell\ell}$ ${\raisebox{-.6ex}{\rlap{$\,\sim\,$}} \raisebox{.4ex}{$\,<\,$}} $ 400 GeV. 

Furthermore, as previously mentioned, the reconstructed $A_{\rm FB}^*$ experiences a partial cancellation of systematical uncertainties.
As the uncorrelated and correlated components of the uncertainty are determined for $A_{\rm FB}^*$ and added up in quadrature, the result is that the global error is significantly reduced
compared to the case of the pure differential cross sections.
From this point of view, the measured $A_{\rm FB}^*$ could be highly competitive with the single-differential cross section and even with the recently measured triple-differential cross section~\cite{Aaboud:2017ffb}
from which it can be obtained. 

We here consider the differential distribution of the $A_{\rm FB}^*$ observable in the di-electron final state, focusing on the invariant mass region around the $Z$-boson pole,
where the high statistics due to its resonant production allows very precise experimental measurements.
The results are presented taking into account the resolution and the experimental acceptance and efficiency factors of this channel~\cite{Khachatryan:2014fba}.
When computing the statistical error on the observable, we include the QCD next-to-next-to-leading order (NNLO) corrections to the cross section~\cite{Hamberg:1990np, Harlander:2002wh} to estimate the expected number of events.    We neglect  
the additional NLO EW corrections, which are estimated~\cite{Aaboud:2017ffb}  to 
give rise to a K-factor of about 1.035 in the di-lepton mass range considered.  

Following up on the results we already presented in~\cite{Accomando:2017scx}, here we further explore the effects of imposing a rapidity cut on the di-lepton system for the observable $A_{\rm FB}^*$.
As it is well known from the literature, such a kinematical cut improves the efficiency of the reconstruction so that a closer shape to the original $A_{\rm FB}$ is recovered.
Conversely, as we mentioned before for the CC channel, the discrepancy in the $A_{\rm FB}^*$ predictions obtained with any two different PDF sets are enhanced by the effect of the rapidity cut.
We extend here the results of~\cite{Accomando:2017scx} by including additional PDF sets.

\begin{figure}[h]
\begin{center}
\includegraphics[width=0.47\textwidth]{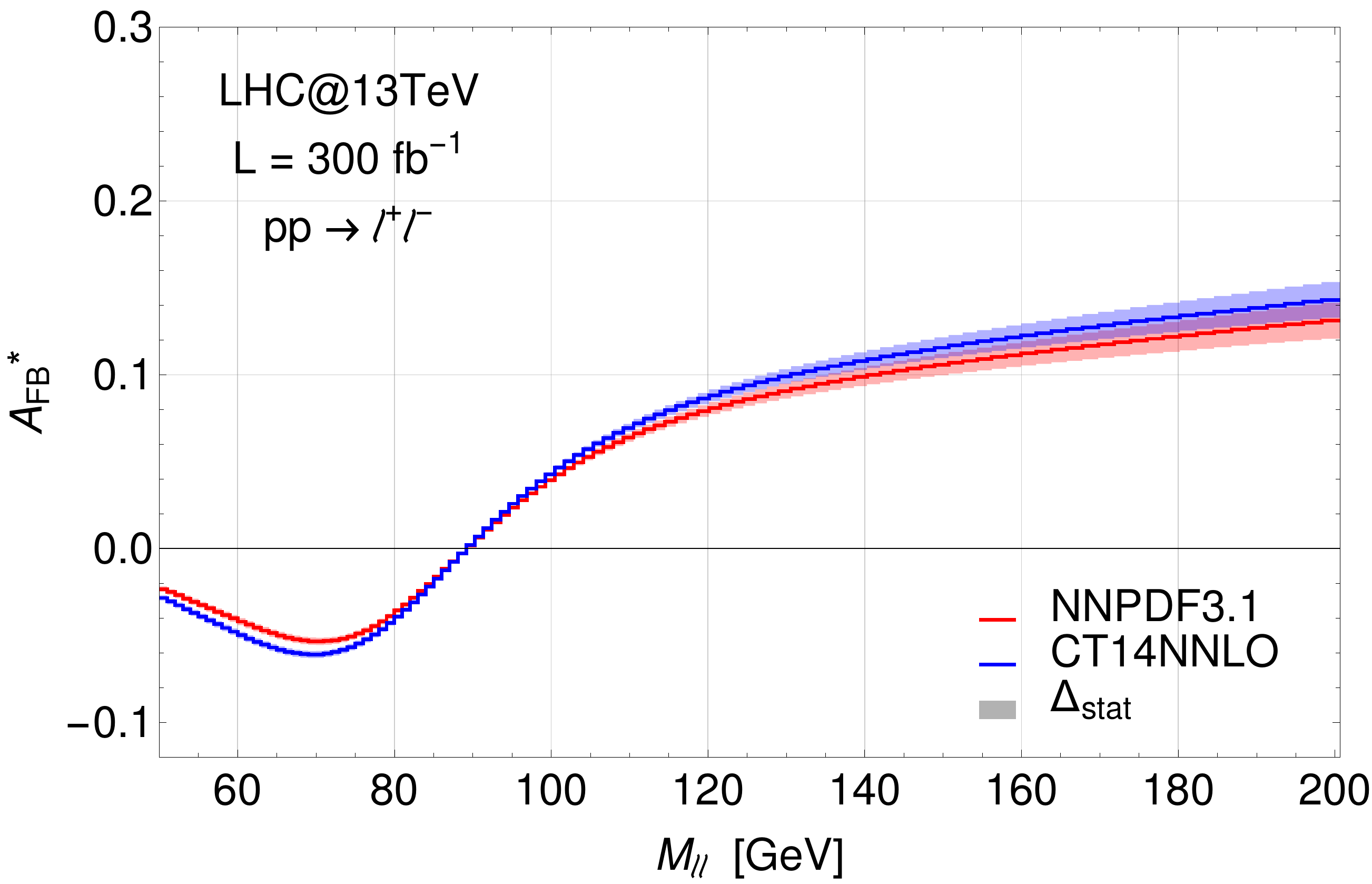}{(a)}
\includegraphics[width=0.47\textwidth]{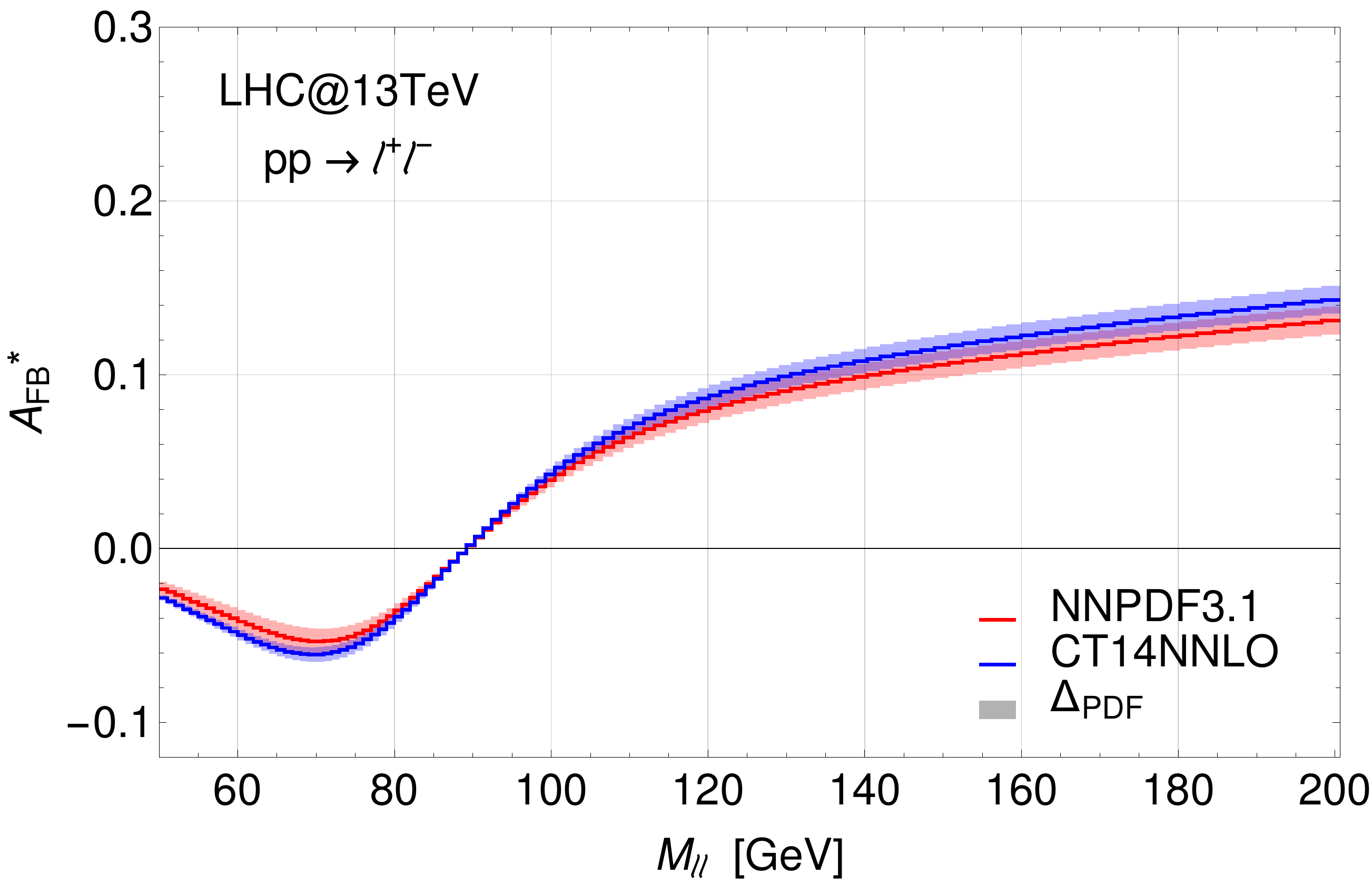}{(b)}
\includegraphics[width=0.47\textwidth]{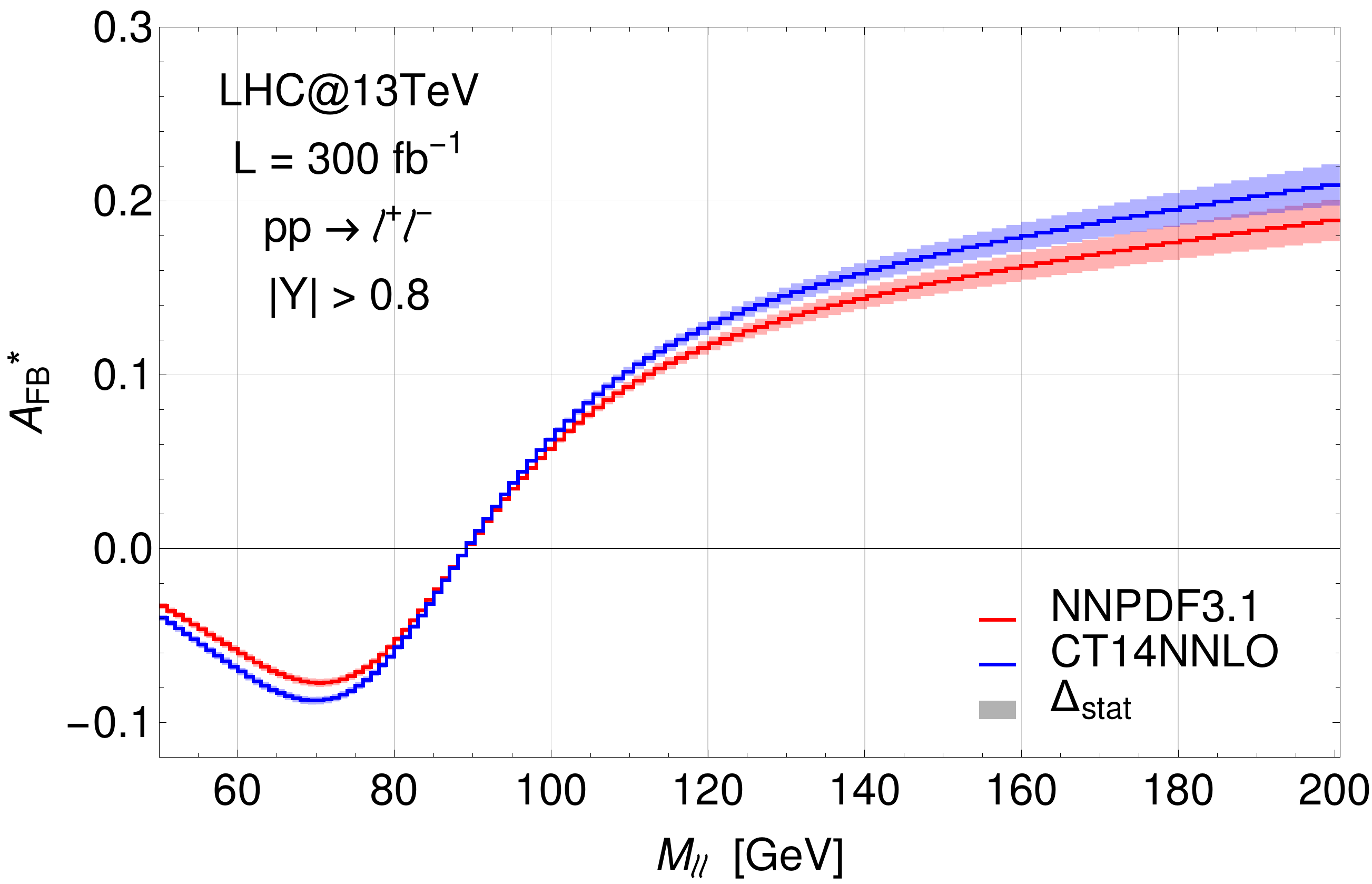}{(c)}
\includegraphics[width=0.47\textwidth]{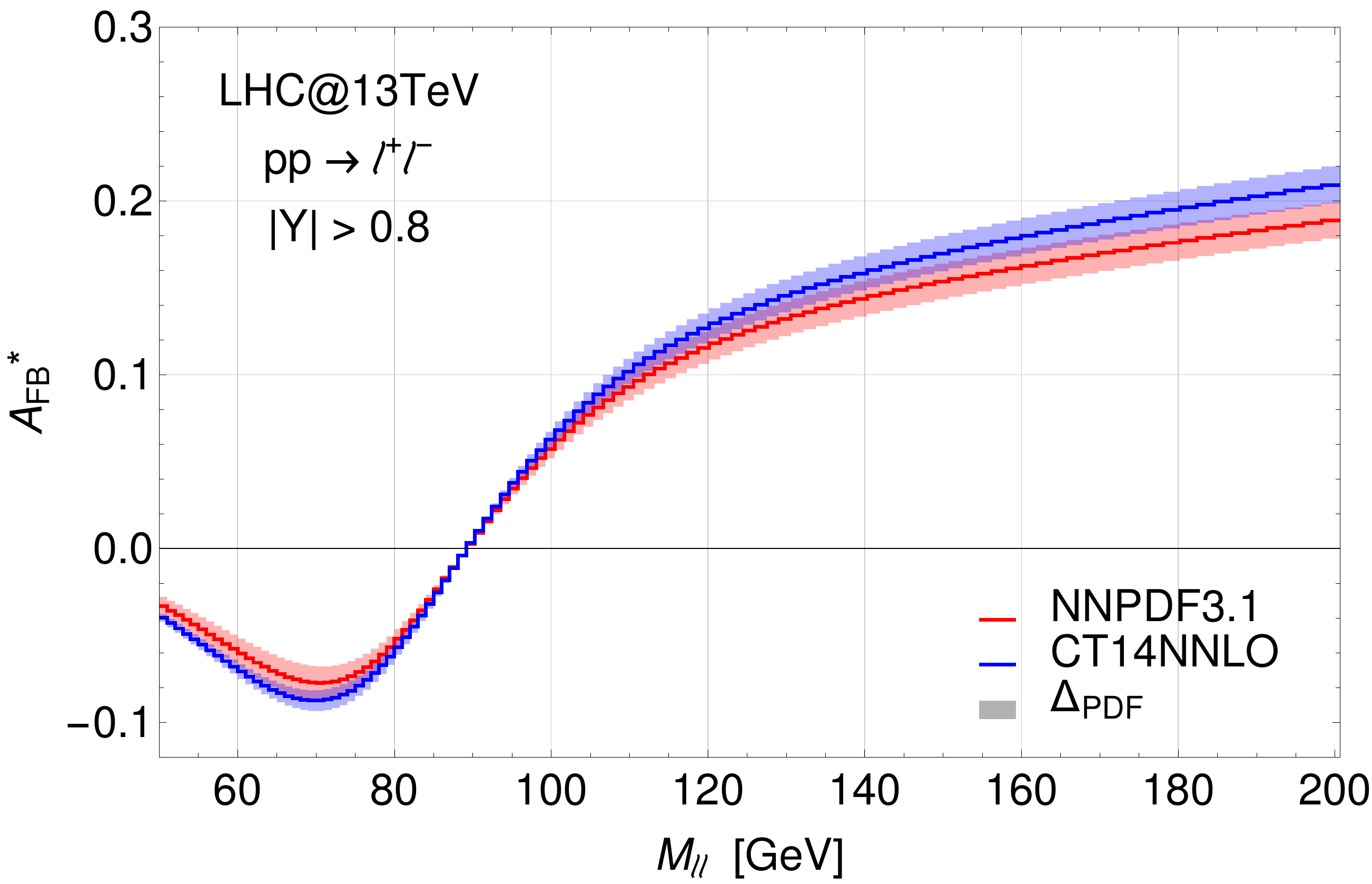}{(d)}
\includegraphics[width=0.47\textwidth]{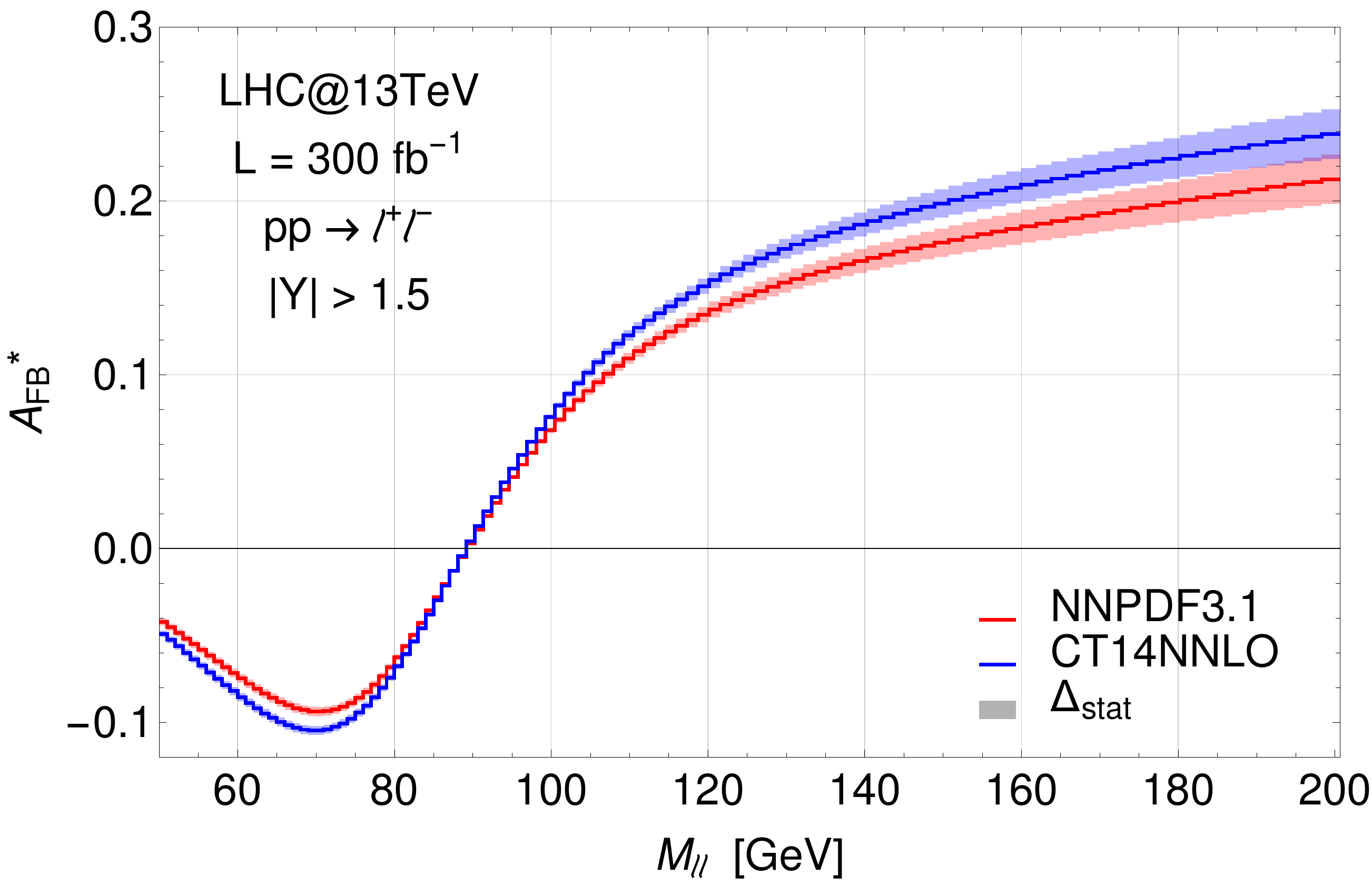}{(e)}
\includegraphics[width=0.47\textwidth]{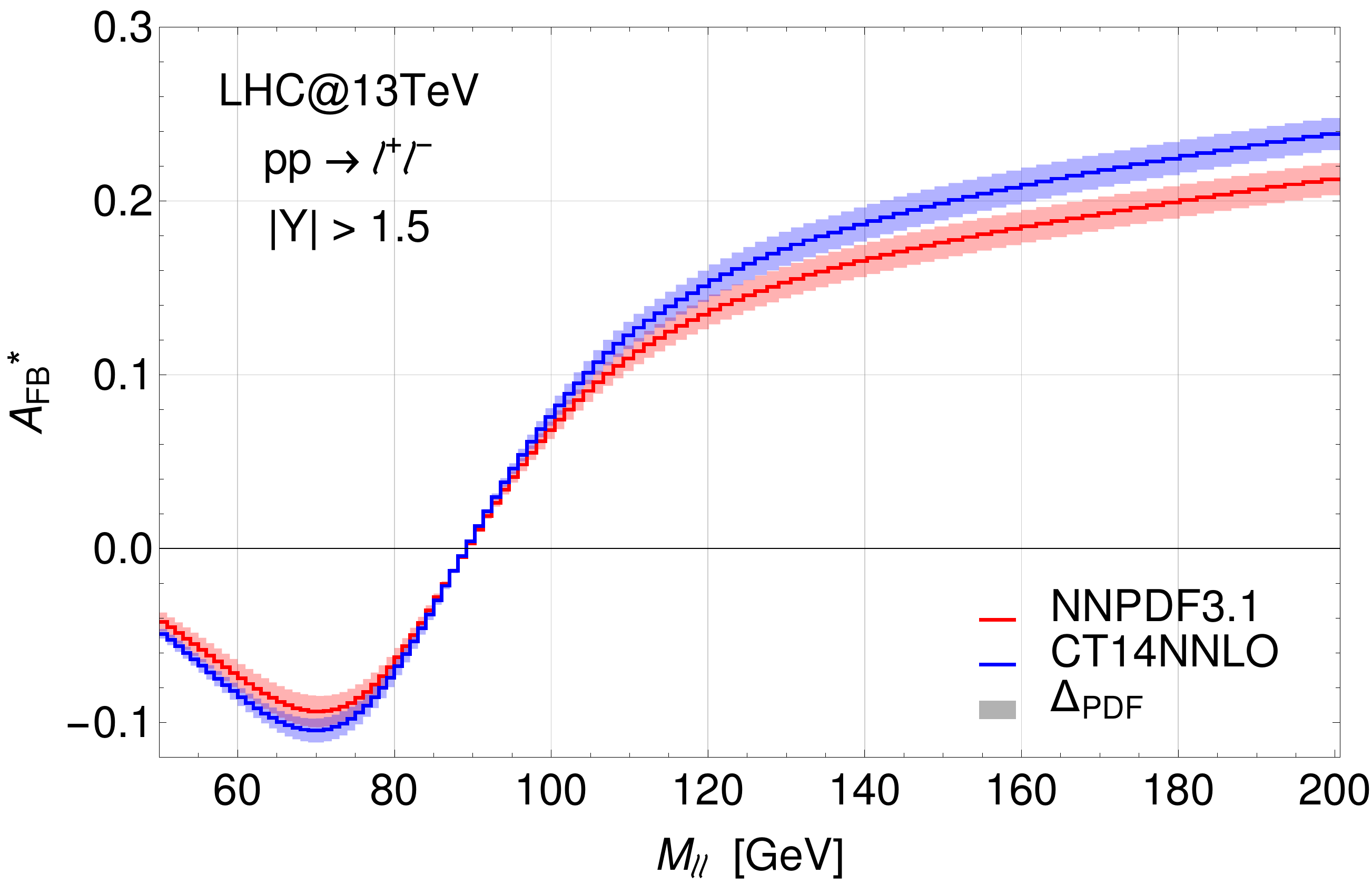}{(f)}
\caption{$A_{\rm FB}^*$ with its statistical (left column) and PDF (right column) error bands obtained with the CT14, and NNPDF3.1 PDF sets
with a rapidity cut of $|y|>0$ (first row), $|y|>0.8$ (second row) and $|y|>1.5$ (third row).}
\label{fig:AFB_y_cut_CT_NNPDF}
\end{center}
\end{figure}

\begin{figure}[h]
\begin{center}
\includegraphics[width=0.47\textwidth]{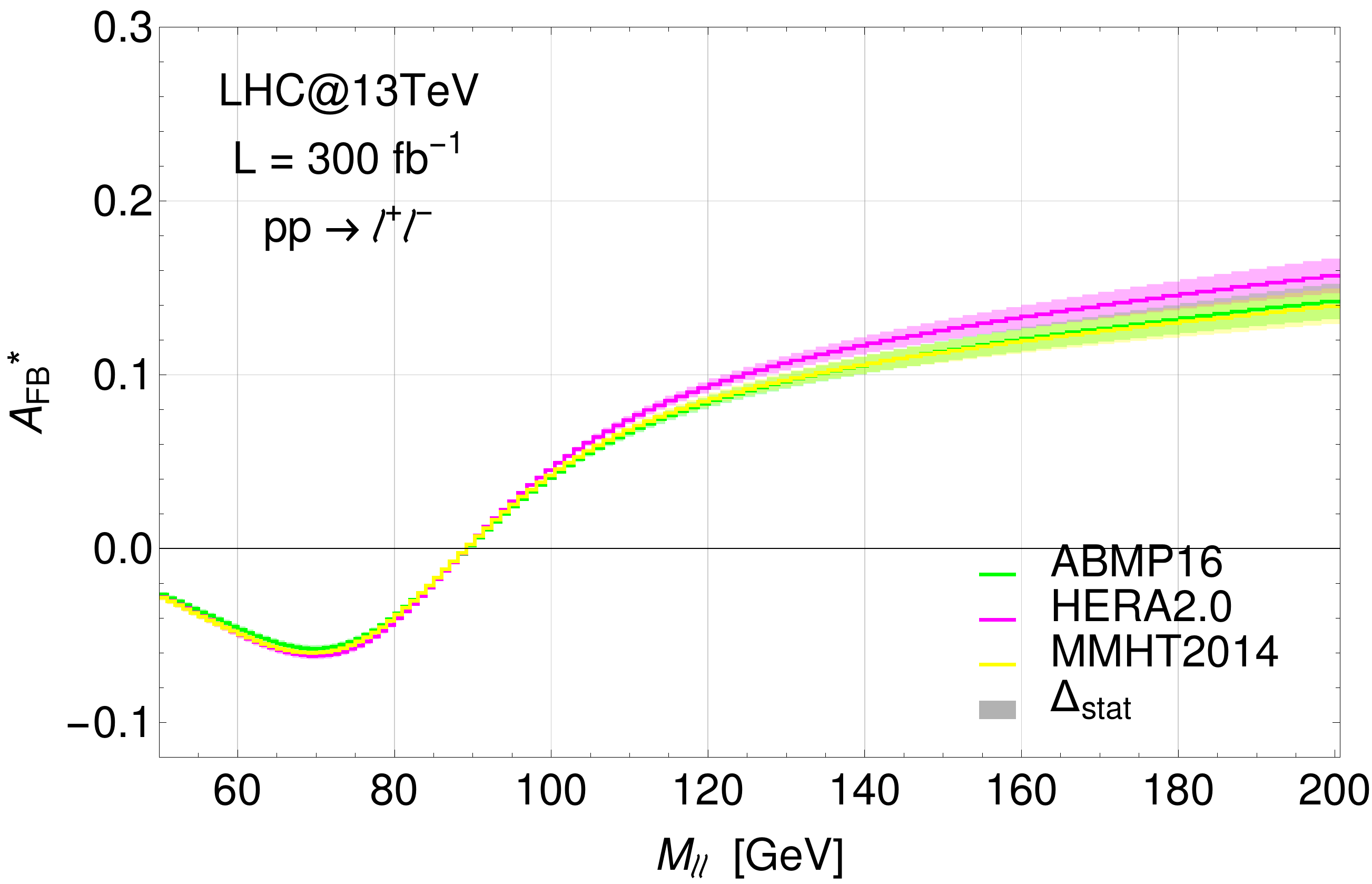}{(a)}
\includegraphics[width=0.47\textwidth]{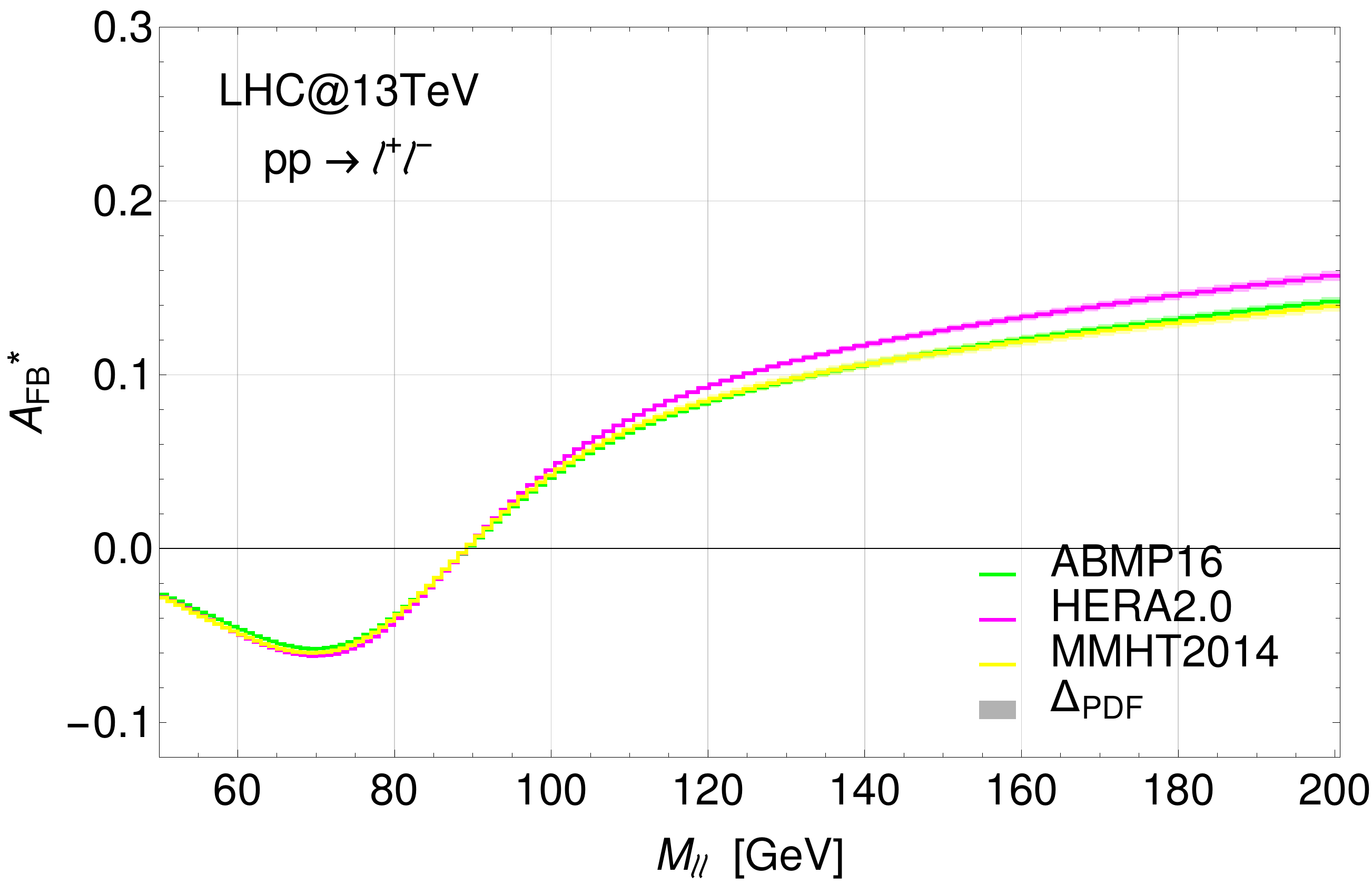}{(b)}
\includegraphics[width=0.47\textwidth]{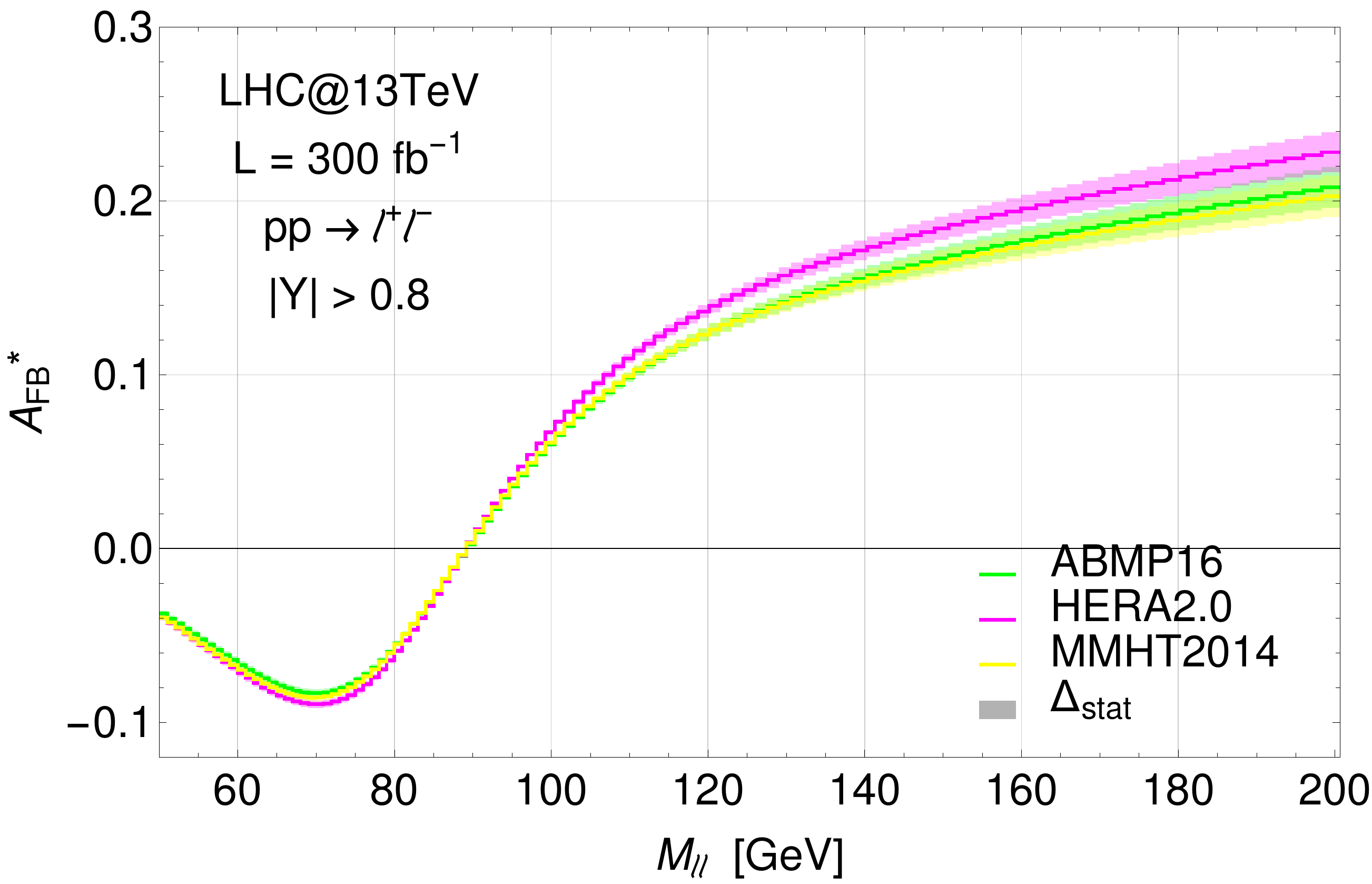}{(c)}
\includegraphics[width=0.47\textwidth]{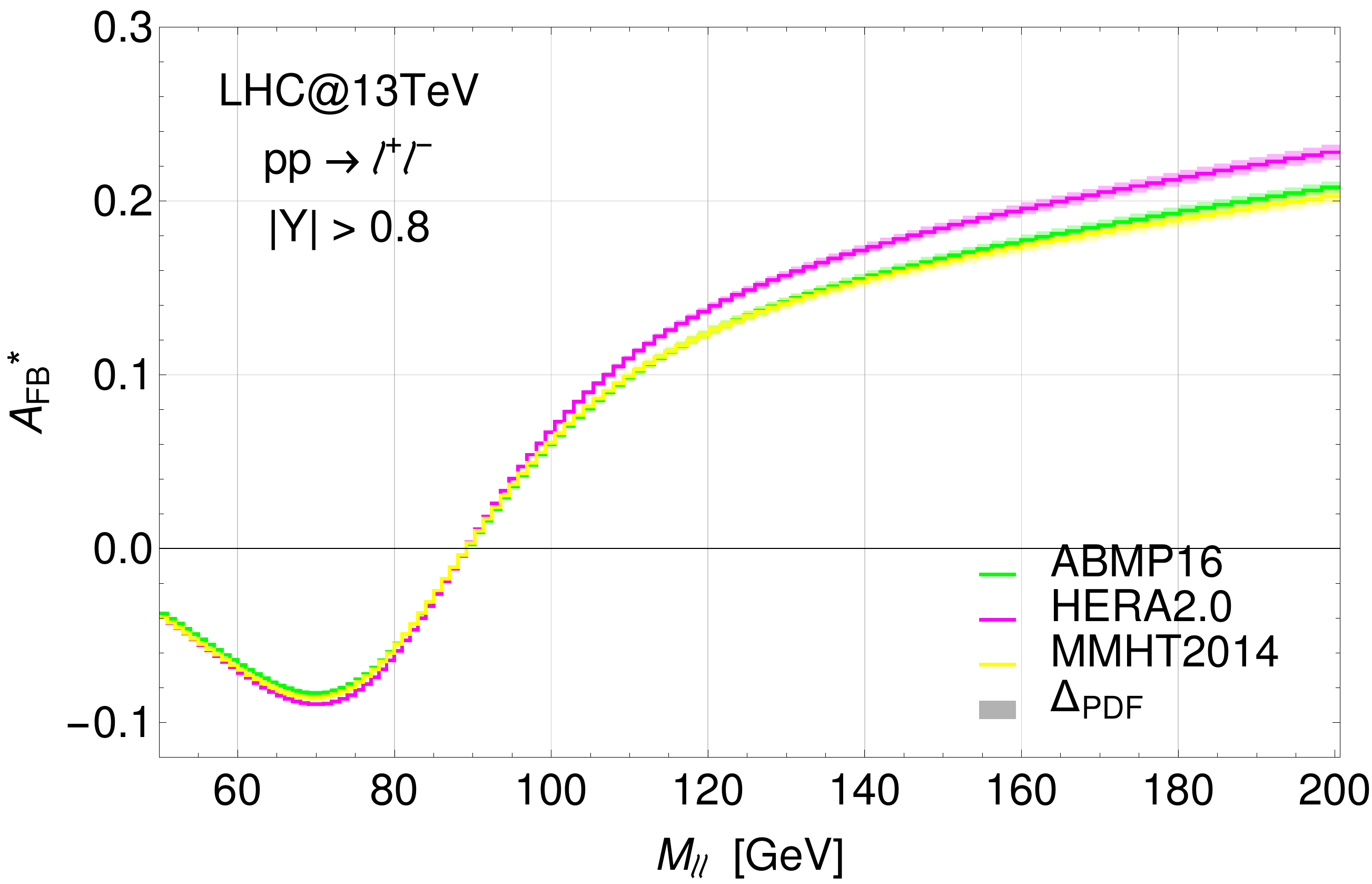}{(d)}
\includegraphics[width=0.47\textwidth]{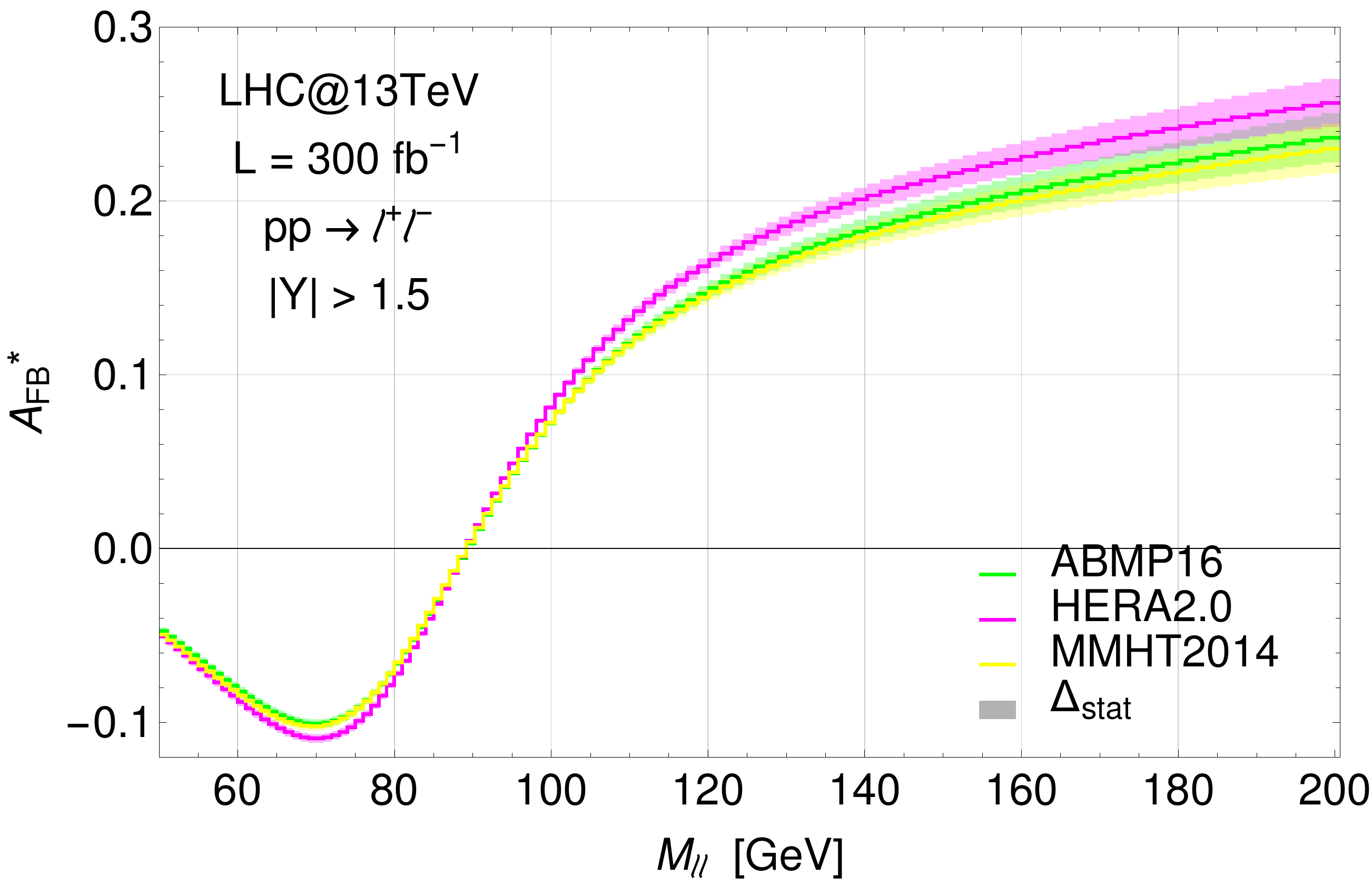}{(e)}
\includegraphics[width=0.47\textwidth]{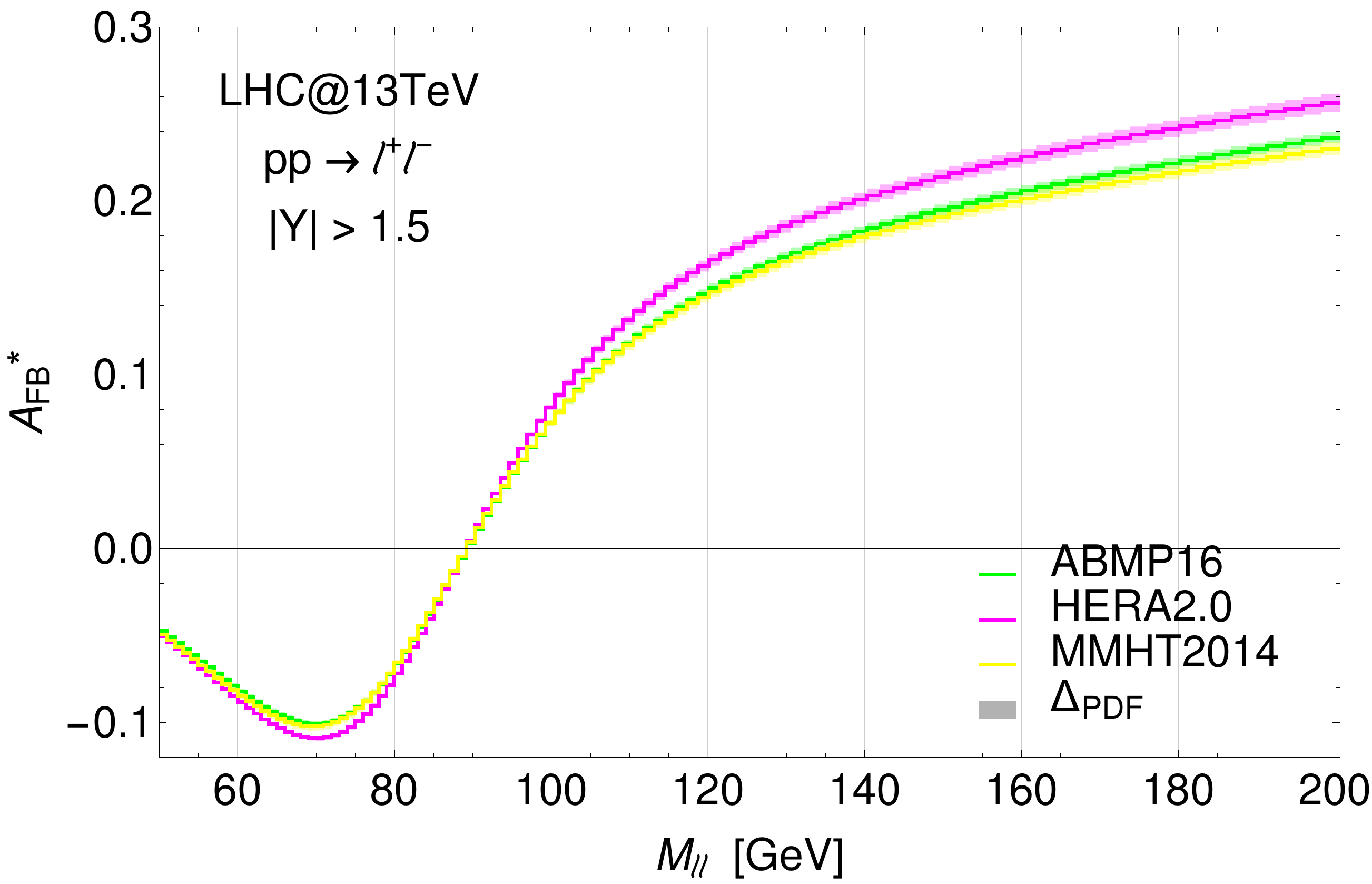}{(f)}
\caption{$A_{\rm FB}^*$ with its statistical (left column) and PDF (right column) error bands obtained with the ABMP16, HERA2.0 and MMHT2014 PDF sets
with a rapidity cut of $|y|>0$ (first row), $|y|>0.8$ (second row) and $|y|>1.5$ (third row).}
\label{fig:AFB_y_cut_ABMP_HERA_MMHT}
\end{center}
\end{figure}

In Fig.~\ref{fig:AFB_y_cut_CT_NNPDF} and Fig.~\ref{fig:AFB_y_cut_ABMP_HERA_MMHT}, the $A_{\rm FB}^*$ distributions in the di-lepton invariant mass have been obtained with different PDF sets:
CT14nnlo~\cite{Dulat:2015mca} and NNPDF3.1nnlo~\cite{Ball:2017nwa} in the former, ABMP16nnlo~\cite{Alekhin:2017kpj}, HERA2.0nnlo~\cite{Abramowicz:2015mha} and MMHT2014nnlo~\cite{Harland-Lang:2014zoa} in the latter.
We have imposed a lower rapidity cut of $|y|>0$ (first row), $|y|>0.8$ (second row) and $|y|>1.5$ (third row) on the di-electron final state.
The distributions are accompanied by the PDF (plots on the right) and the statistical (plots on the left) error bands.
The latter have been computed assuming an integrated luminosity of 300 fb$^{-1}$, that is, the value which is supposed to be achieved by the end of the LHC Run-II programme.

The increasing discrepancies between the predictions as the rapidity cut gets stronger are a consequence of the different parameterisations in the partonic content of the proton, characteristic of each PDF’s set.
Quarks and anti-quarks PDFs suffer from larger uncertainties in the regions of high-$x$ and low-$x$, respectively: the rapidity cut indeed selects these kinematical regions.
Future experimental measurements of $A_{\rm FB}^*$ in these invariant mass and high rapidity regions have the potential of reaching sufficient precision to discriminate between different PDF parameterisations in the corresponding $x$ range.
If using $A_{\rm FB}^*$, which is a cross section ratio, this discrimination can be remarkably outside the PDF’s uncertainty bands, as shown by the right-hand side plots.
This is true in the invariant mass interval just before the $Z$-boson peak, 60 GeV$\le M(ll)\le$ 80 GeV, and soon after it, 100 GeV$\le M(ll)\le$ 180 GeV, for a high enough $y(ll)$ cut.
This condition is not at all guaranteed by the triple differential cross section that contains the same information  but is not normalised.
This powerful feature of the $A_{\rm FB}^*$ observable is not taken into account, generally, and it should.

From Fig.~\ref{fig:AFB_y_cut_ABMP_HERA_MMHT}, one can clearly distinguish the HERA2.0 predictions from the ABMP16 and MMHT2014 ones already for $y(ll)\ge 0.8$.
Analogous results are evident in Fig.~\ref{fig:AFB_y_cut_CT_NNPDF} in the comparison between the NNPDF3.1 and the CT14NNLO predictions and they were already displayed by the authors in Ref.~\cite{Accomando:2017scx}.

\section{Parton luminosities at high rapidity}
\label{sec:parton_luminosities}

\begin{figure}[h]
\begin{center}
\includegraphics[width=0.47\textwidth]{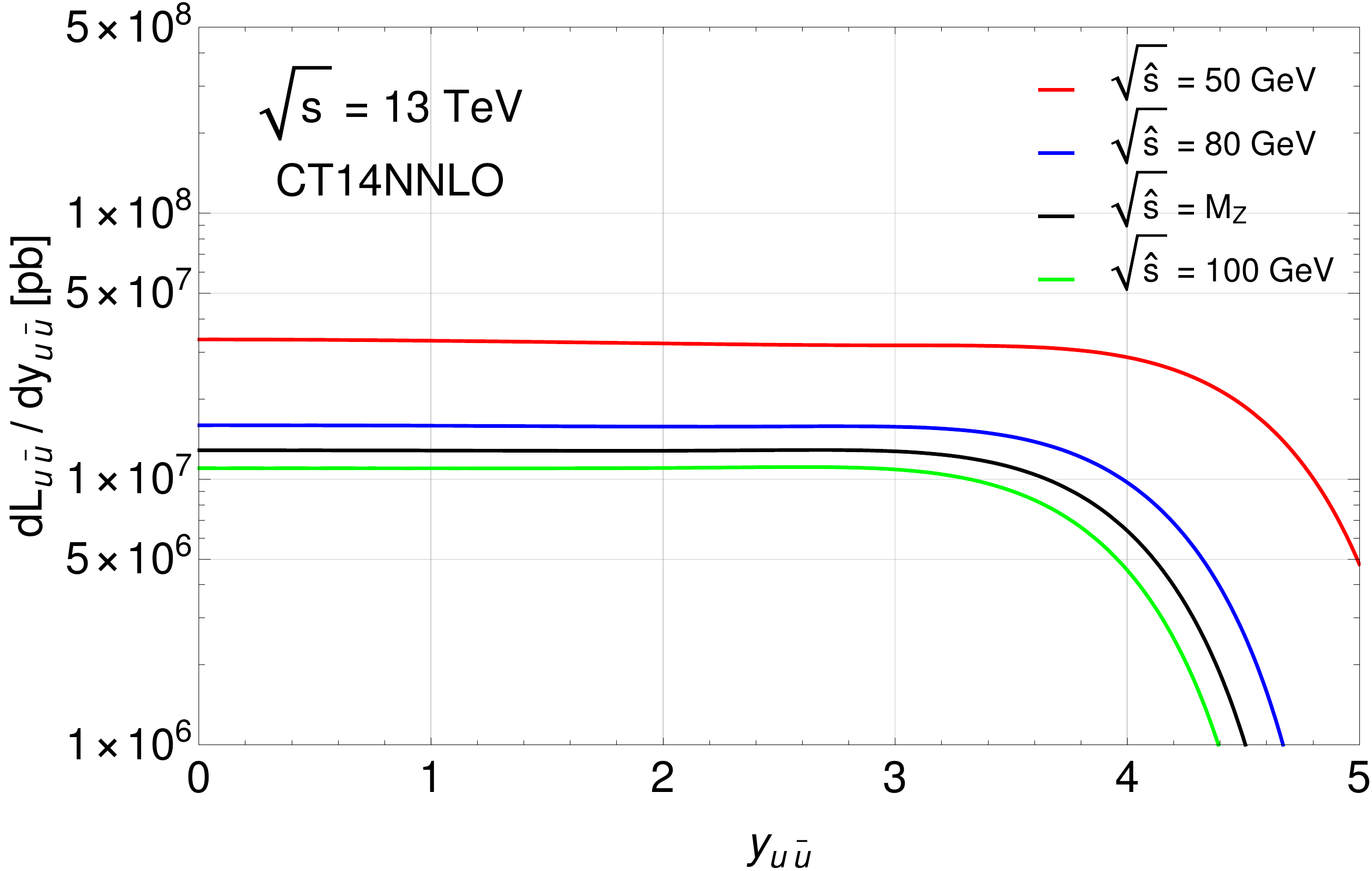}{(a)}
\includegraphics[width=0.47\textwidth]{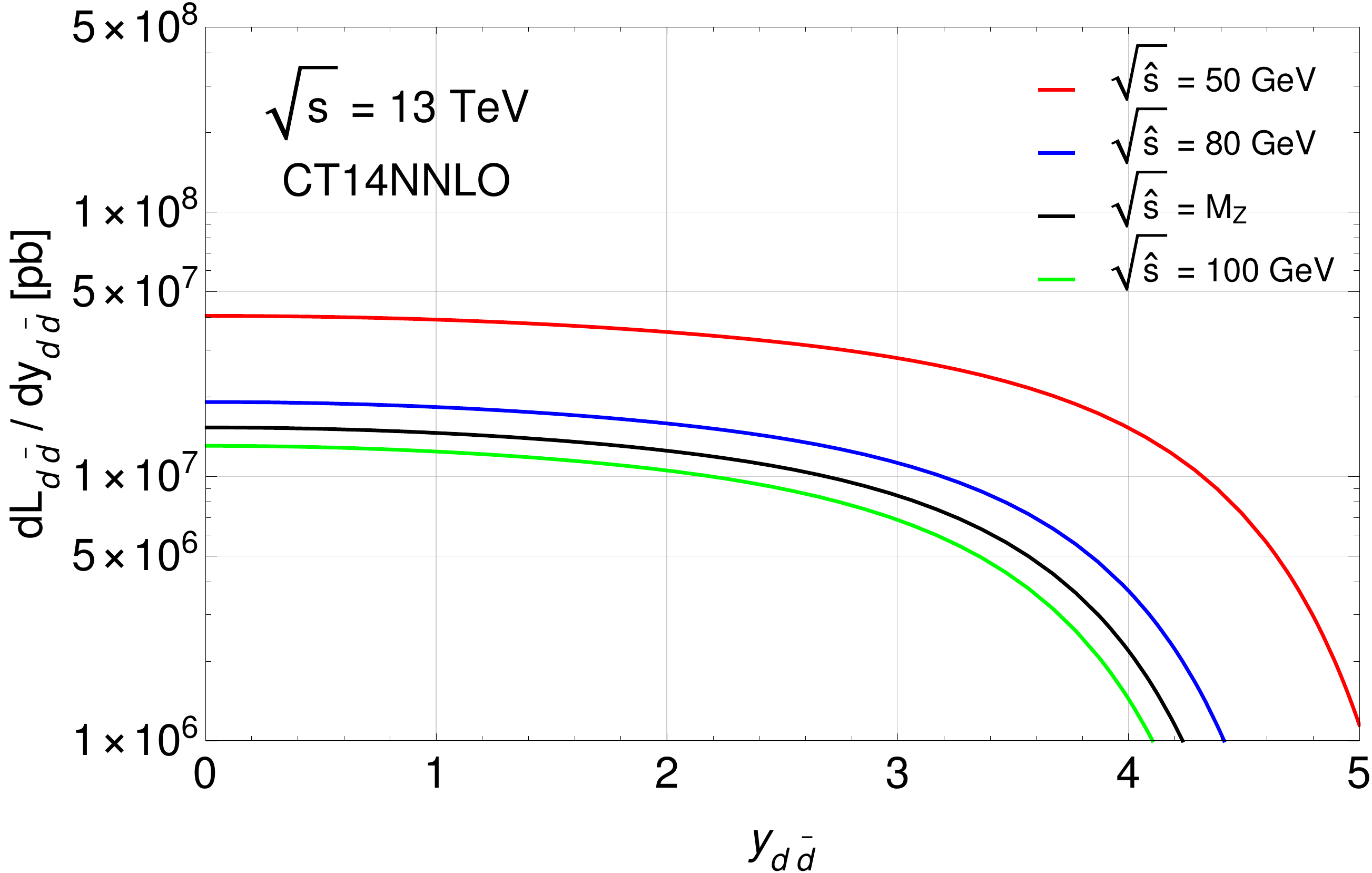}{(b)}
\caption{Parton luminosities for the (a) $u-\bar{u}$ and (b) $d-\bar{d}$ interactions as a function of the partonic c.o.m. rapidity, computed at four different c.o.m. energies around the $Z$-boson peak.}
\label{fig:Lum_uu_dd}
\end{center}
\end{figure}

In this section we want to further investigate the origin of the behaviour of $A_{\rm FB}^*$ that we have observed.
We will show that such a behaviour is closely related to the flavour content of the proton and can give access to the up-type PDFs in a unique way.
To this aim, we compare the parton luminosities from the $u-\bar{u}$ and the $d-\bar{d}$ interactions at some fixed value of the centre-of-mass energy around the $Z$-boson peak
focusing on their dependence on the partonic rapidity, $y_{q\bar q}$.
We follow the formulae of Ref.~\cite{Quigg:2009gg} and in Fig.~\ref{fig:Lum_uu_dd} we show the 
predictions for the CT14 set. 

\begin{figure}[h]
\begin{center}
\includegraphics[width=0.47\textwidth]{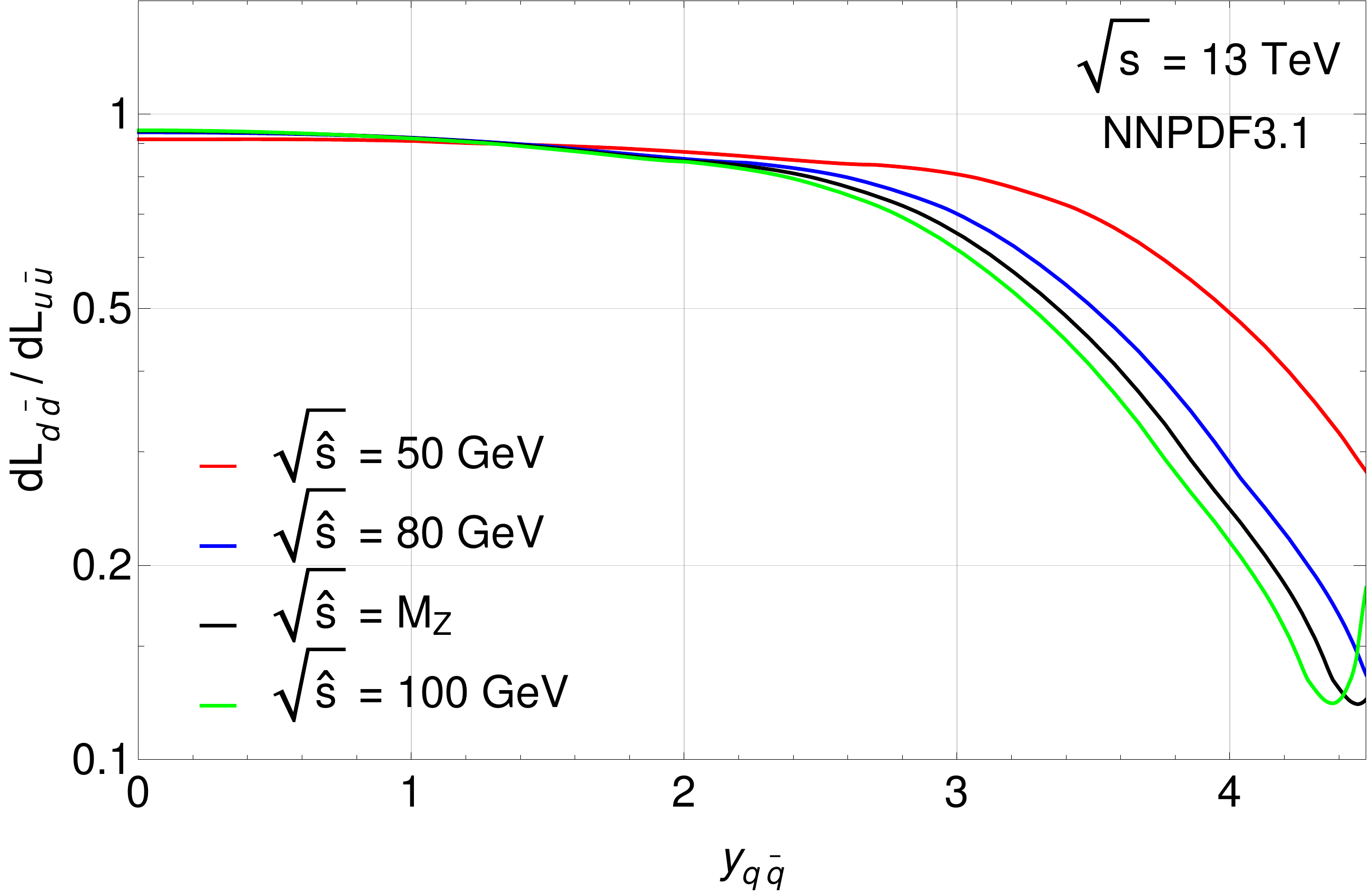}{(a)}
\includegraphics[width=0.47\textwidth]{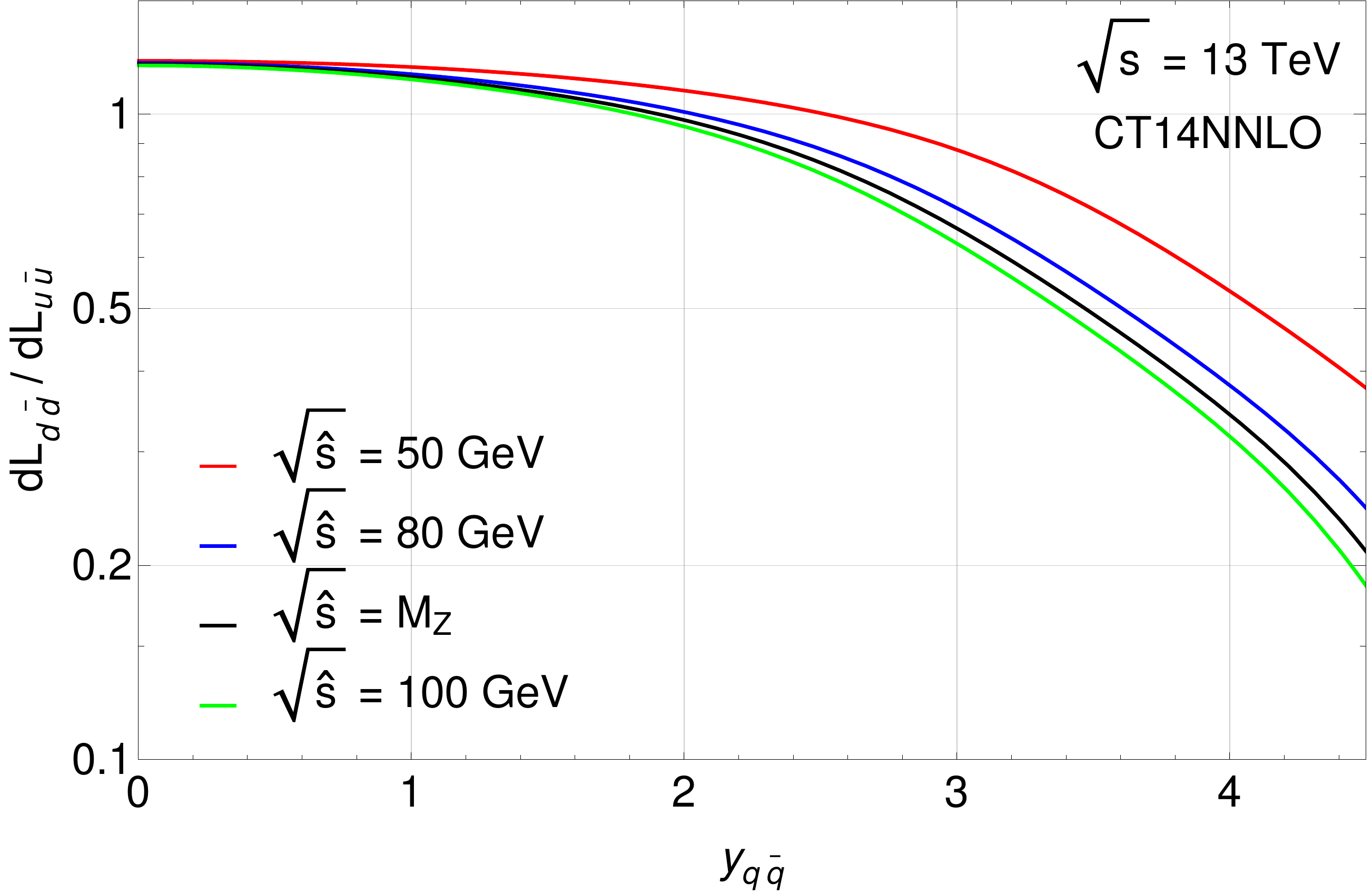}{(b)}
\includegraphics[width=0.47\textwidth]{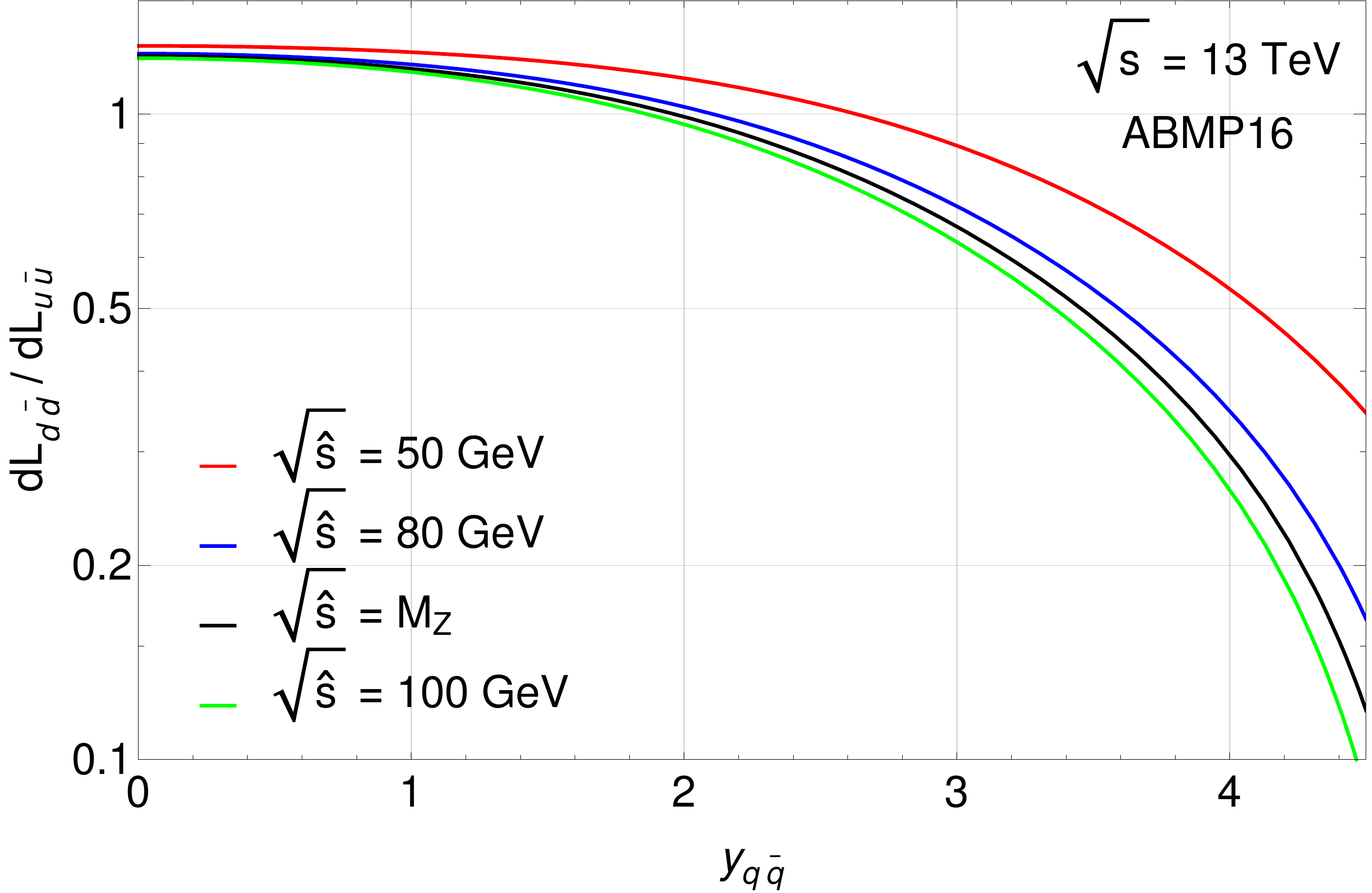}{(c)}
\includegraphics[width=0.47\textwidth]{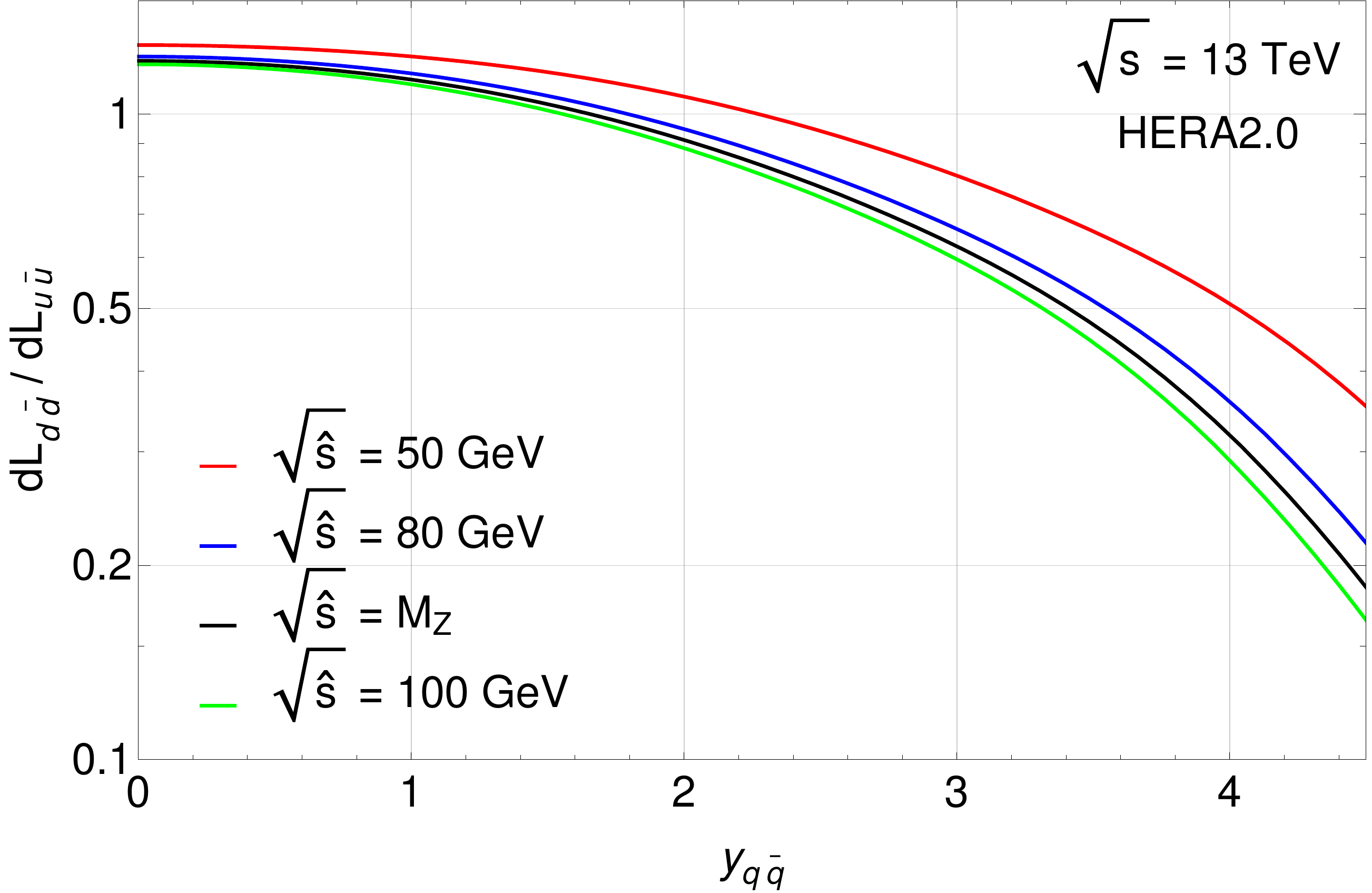}{(d)}
\includegraphics[width=0.47\textwidth]{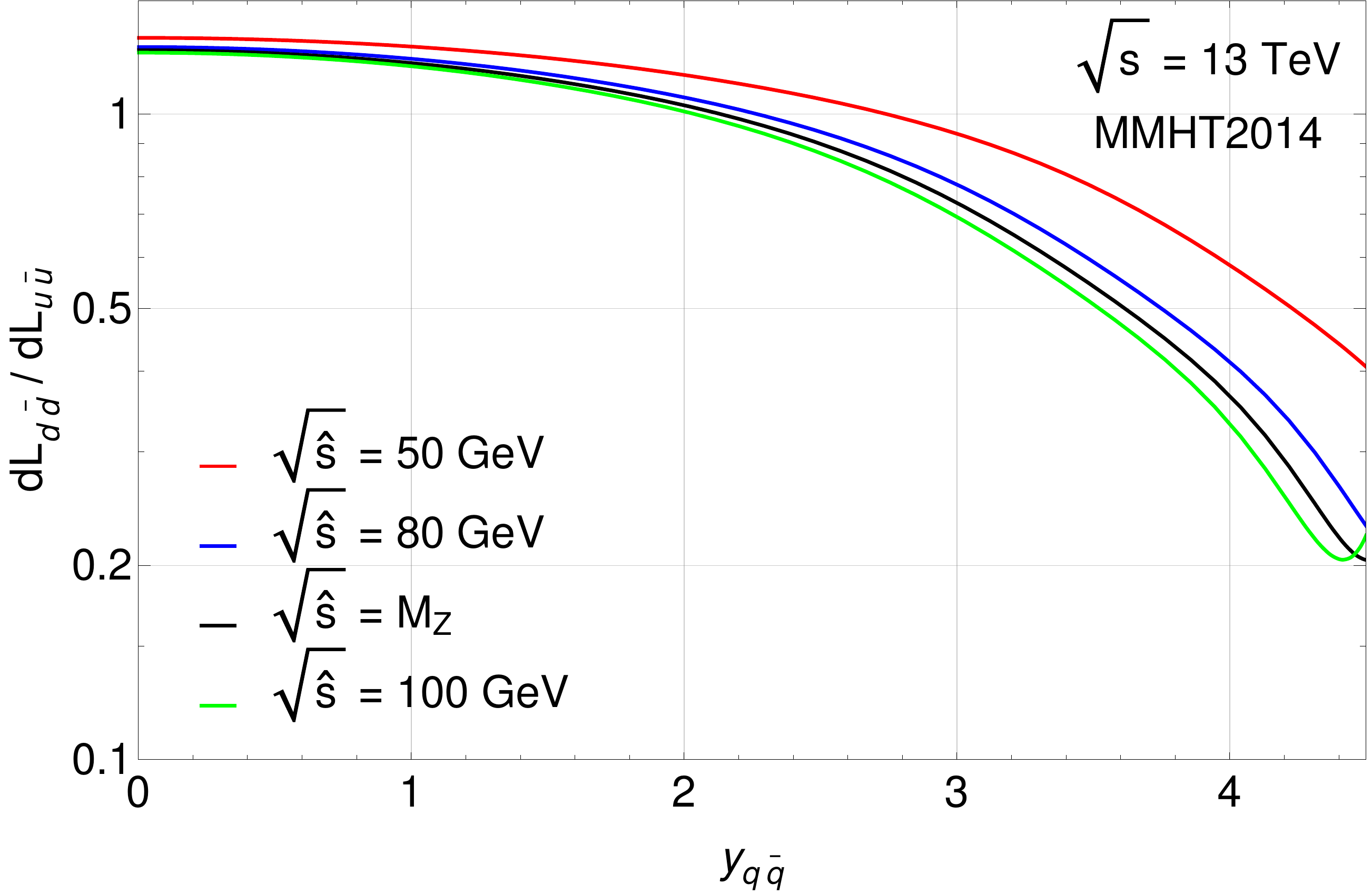}{(e)}
\includegraphics[width=0.47\textwidth]{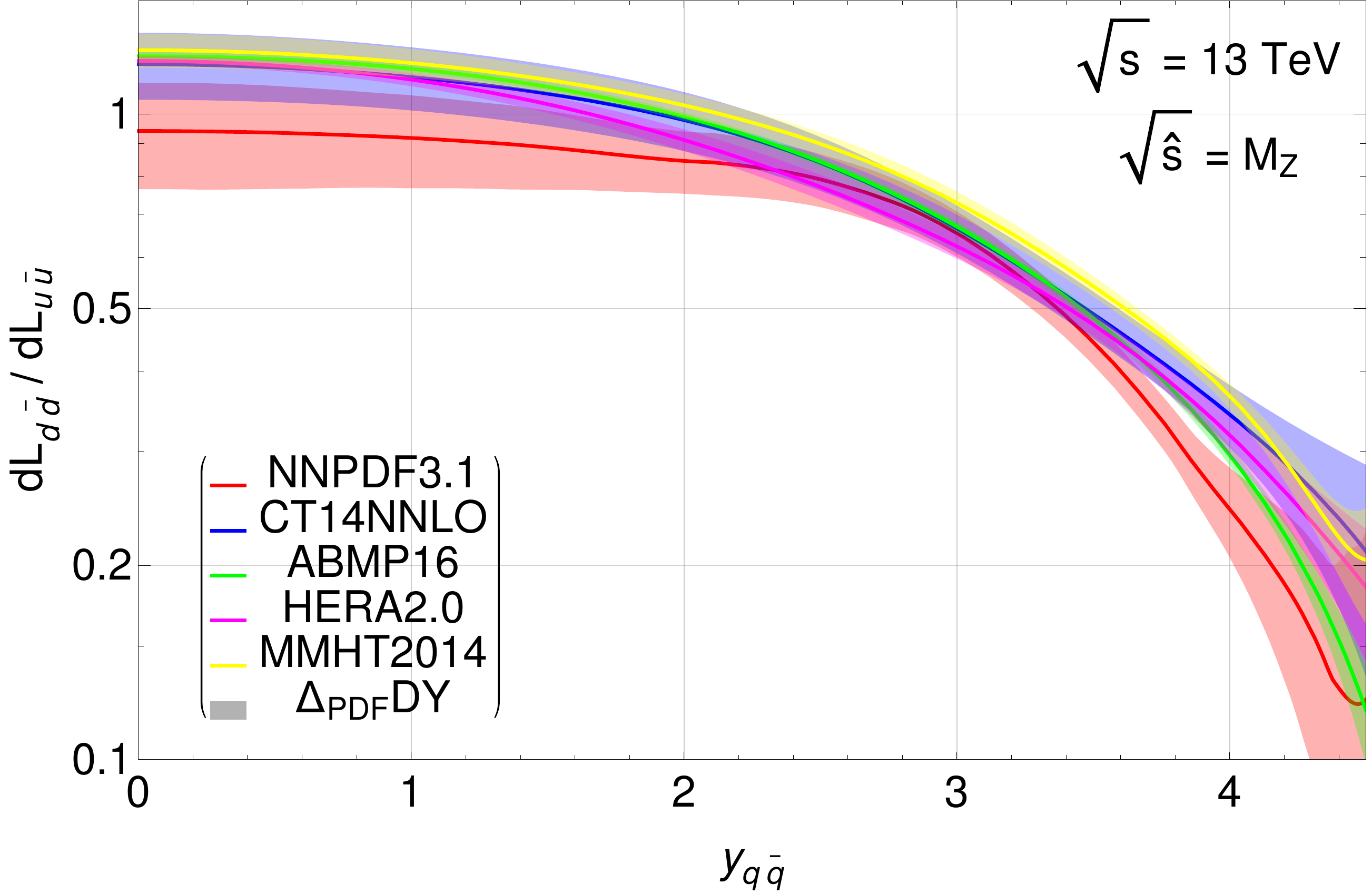}{(f)}
\caption{Ratio of the $d-\bar{d}$ and $u-\bar{u}$ parton luminosities as a function of the c.o.m. rapidity at four fixed values of the c.o.m. energy and for the PDF sets (a) NNPDF3.1, (b) CT14, (c) ABMP16, (d) HERA2.0, (e) MMHT2014.
(f) Same ratio including the PDF error band for all the sets at fixed energy $\sqrt{\hat s} = M_Z$.}
\label{fig:Lum_ratio}
\end{center}
\end{figure}

As visible, the $d-\bar{d}$ initiated process gets more suppressed at high rapidity in comparison to the $u-\bar{u}$ one.
This effect gets enhanced with increasing the centre-of-mass energy of the hard quark-antiquark scattering.
It is indeed particularly evident on the right hand side of the $Z$-boson peak.
It is then interesting to visualise the ratio of the $u-\bar{u}$ and $d-\bar{d}$ partonic luminosities as a function of the c.o.m. rapidity for the various PDF sets that we are considering.
This is shown in Figs.~\ref{fig:Lum_ratio}(a)—(e) for the four values of the centre-of-mass energy of the hard quark-antiquark scattering that, on purpose, have been selected before and after the $Z$-boson peak (see legend).
While at low rapidity, $y_{q\bar q}$, the two luminosities are comparable, with increasing $y_{q\bar q}$ the $d-\bar{d}$ component gets depleted by roughly an order of magnitude.
The level of suppression of the $d-\bar{d}$ contribution is different within each PDF set.
In order to estimate such differences in the predictions, one needs to include the PDF errors from each set.

In Fig.~\ref{fig:Lum_ratio}(f), the ratio of the $u-\bar{u}$ and $d-\bar{d}$ partonic luminosities as a function of the c.o.m. rapidity is displayed with the corresponding PDF error band
as calculated within the various PDF sets at fixed $\sqrt{\hat s} = M_Z$.

In order to detail the displayed differences in the ratio predicted by the various collaborations, we take as reference point the rapidity value $y = 4.5$.
Here, the relative contribution from the down-quark initiated process lies in the interval 2\% - 23\% for the NNPDF3.1 set, 13\% - 29\% for the CT14 set,
10\% - 14\% for the ABMP16, 14\% - 23\% for the HERA2.0 set and 16\% - 25\% for the MMHT2014 set. 

This has two consequences. First of all, there are differences expected in the $Z$-boson production cross sections obtained with the various PDF sets that are well outside the PDF’s error band at large rapidity.
A good sensitivity could be reached to small differences in the $u$, $d$ quark and anti-quark parameterisation.
Second, by imposing a sufficiently high rapidity cut on the data sample, all the considered PDF sets agree within errors that the $d\bar d$ component of the luminosity gets reduced to at most 20$\%$.
One could therefore have a direct handle on the processes that are initiated by the $u-\bar{u}$ interactions.
This implies that one could have direct access to the valence $u$-quark and the sea $u$ quark and anti-quark PDFs.
Up to some degree of purity, the up-quark flavour could be extracted from the neutral Drell-Yan data and the corresponding PDFs constrained (or tuned) by a direct comparison.
Using the valence $u$-quark PDF already well measured, one could adjust the sea $u$ and $\bar u$ distributions to fit the data.
The kinematical region of large rapidity corresponds to selecting 
 a low $x_{\bar u}$ and a high $x_u$ in the $u\bar u$ scattering process.
These are the ranges least known for the quarks PDFs.
Therefore, at large rapidity the neutral Drell-Yan data could be used to constrain the $u$ and $\bar u$ PDFs and determine the corresponding $u\bar u$ parton luminosity in these two extreme regions of $x$. 

\section{Sensitivity to the $u$ and $\bar u$ PDFs}
\label{sec:sensitivity}

\begin{figure}[h]
\begin{center}
\includegraphics[width=0.47\textwidth]{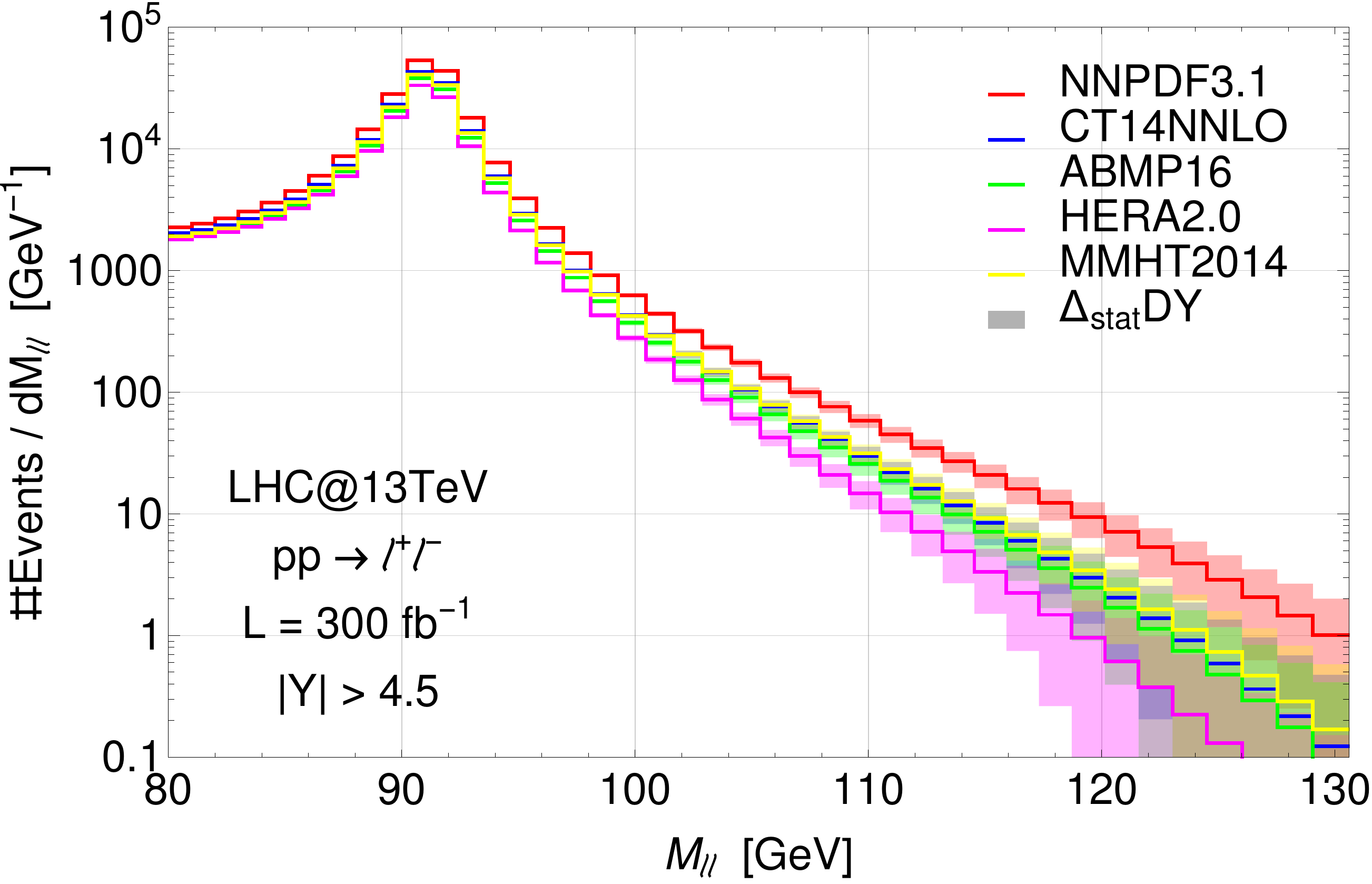}{(a)}
\includegraphics[width=0.47\textwidth]{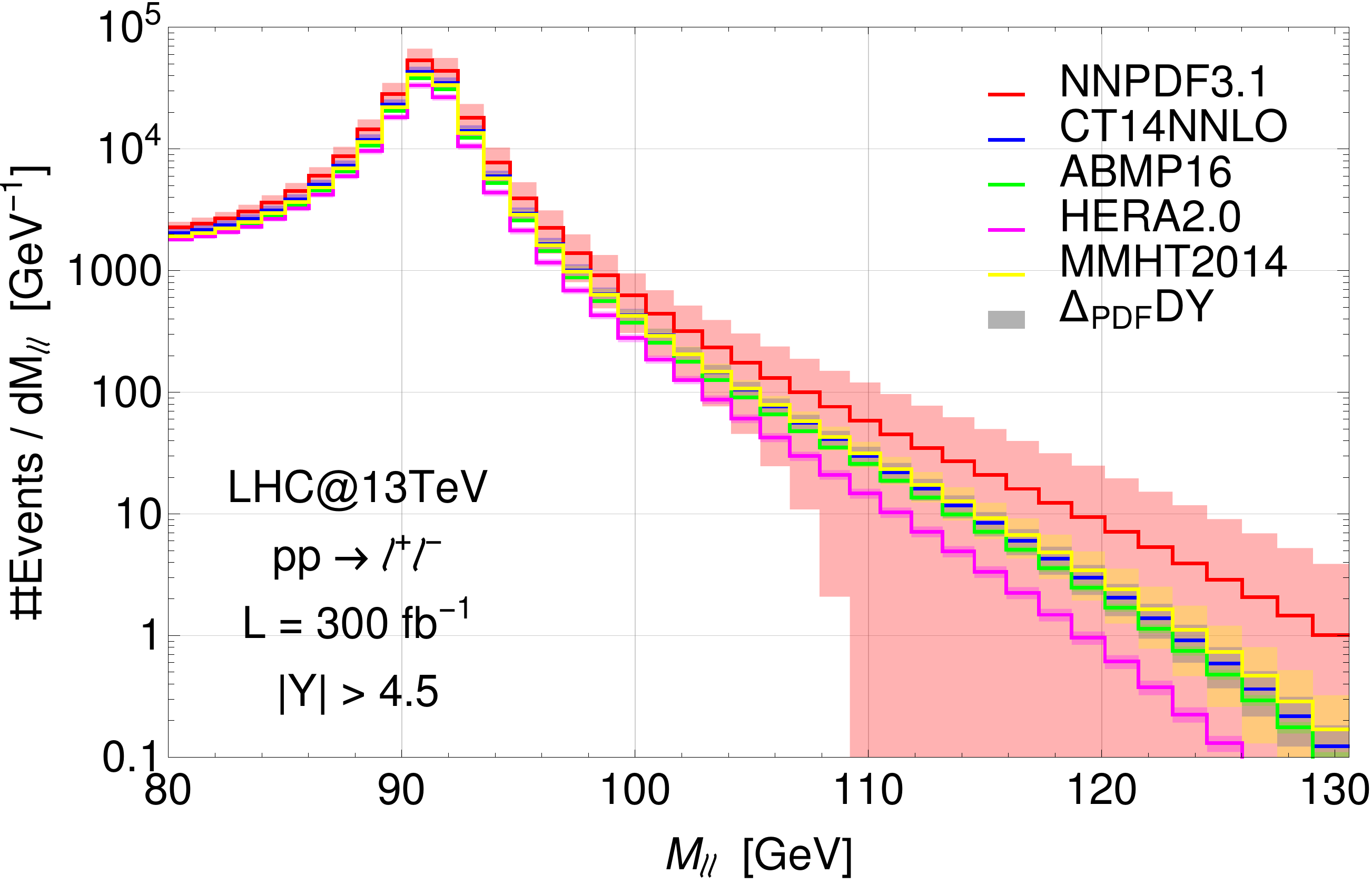}{(b)}
\caption{Number of events as a function of the di-lepton invariant mass including the (a) statistical and (b) PDF errors for a rapidity cut $|Y_{ll}| > 4.5$. A luminosity $L = 300$ fb$^{-1}$ is assumed and acceptance cuts $|\eta| < 5$ and $p_T > 20$ GeV are imposed on both leptons.}
\label{fig:XS_high_rapidity}
\end{center}
\end{figure}

We now want to exploit this idea as far as the LHC sensitivity would allow. For a meaningful analysis we cannot push the rapidity cut too high due to the reduction of the data sample though.
Moreover in order to explore high di-lepton rapidities, we need to extend the acceptance region of the detector to $|\eta| < 5$.
Experimentally this is possible in the di-electron channel, but the analysis still requires at least one lepton falling into the usual acceptance region $|\eta| < 2.5$~\cite{Aaboud:2017ffb}.
Here we relax this assumption and impose instead a symmetric cut $|\eta| < 5$ on both leptons.
In Fig.~\ref{fig:XS_high_rapidity}, we show the expected number of events as a function of the di-lepton invariant mass obtained for different PDF choices when imposing a rapidity cut $|y| > 4.5$.
For this value, the measurable region would correspond to $x_1\simeq 0.6$ and $x_2\simeq 7\cdot 10^{-4}$ (low-$x$ anti-quark and high-$x$ quark).
This range could be used to constrain the valence/sea up-quarks and sea up-type antiquarks PDFs, where they are least known, and determine the corresponding parton luminosities. 
In the plot on the left we can observe the statistical error band evaluated for an integrated luminosity of 300 fb$^{-1}$.
As visible, in the region around the $Z$-boson peak the statistics is very high despite the stiff kinematical cut, however, the distribution rapidly falls moving away from the resonant production.
Up to $M_{ll}\le 120$ GeV, the various PDF sets could be distinguished if relying only on the statistical error.
Conversely, the plot on the right-hand side shows that with the current PDF uncertainties the predictions obtained with the various PDF sets cannot be disentangled.
They are indeed all within the errors. The spectrum does not seem to be able to constrain the PDF of the valence/sea up-quark and the sea up-type antiquark.

\begin{figure}[h]
\begin{center}
\includegraphics[width=0.47\textwidth]{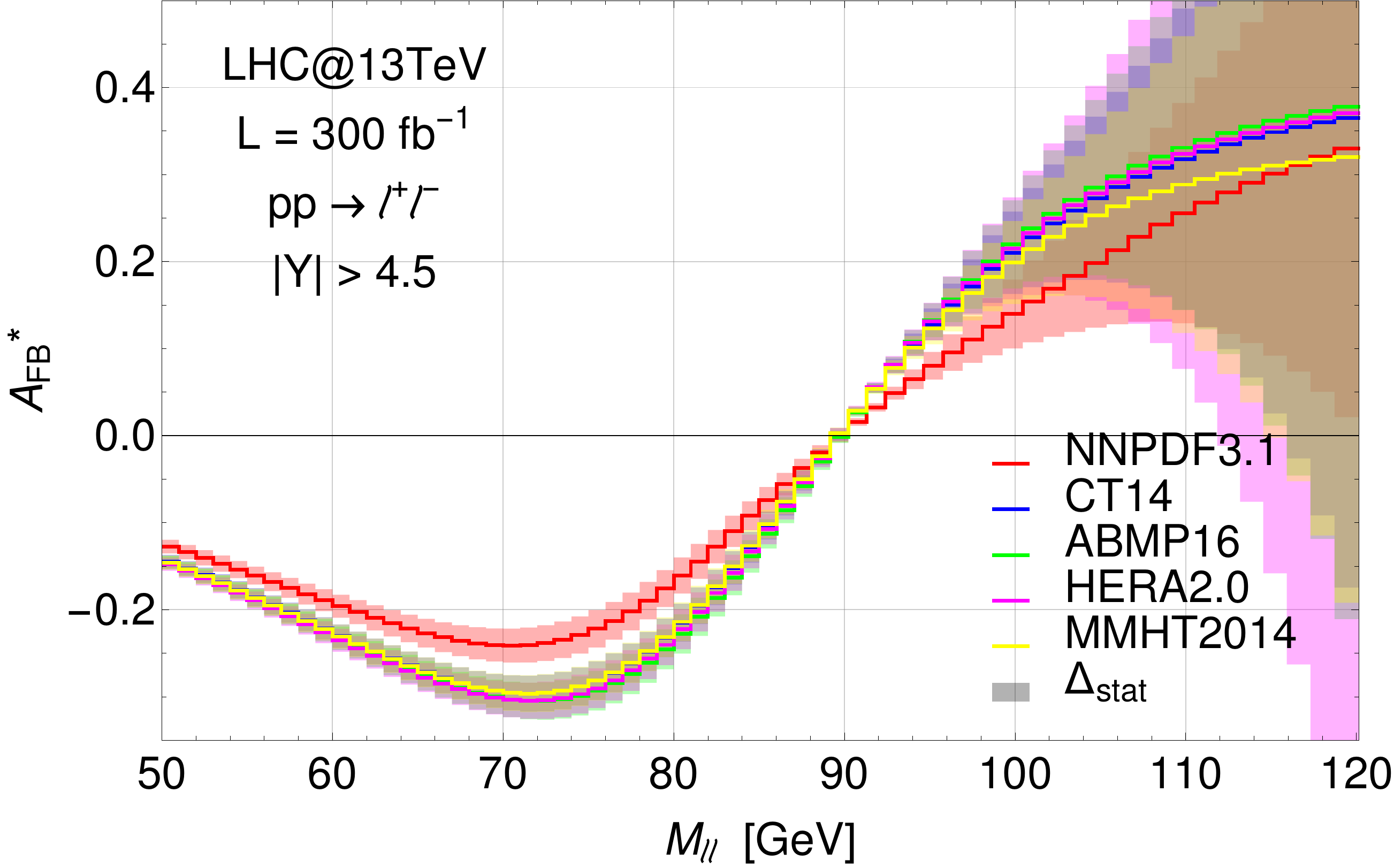}{(a)}
\includegraphics[width=0.47\textwidth]{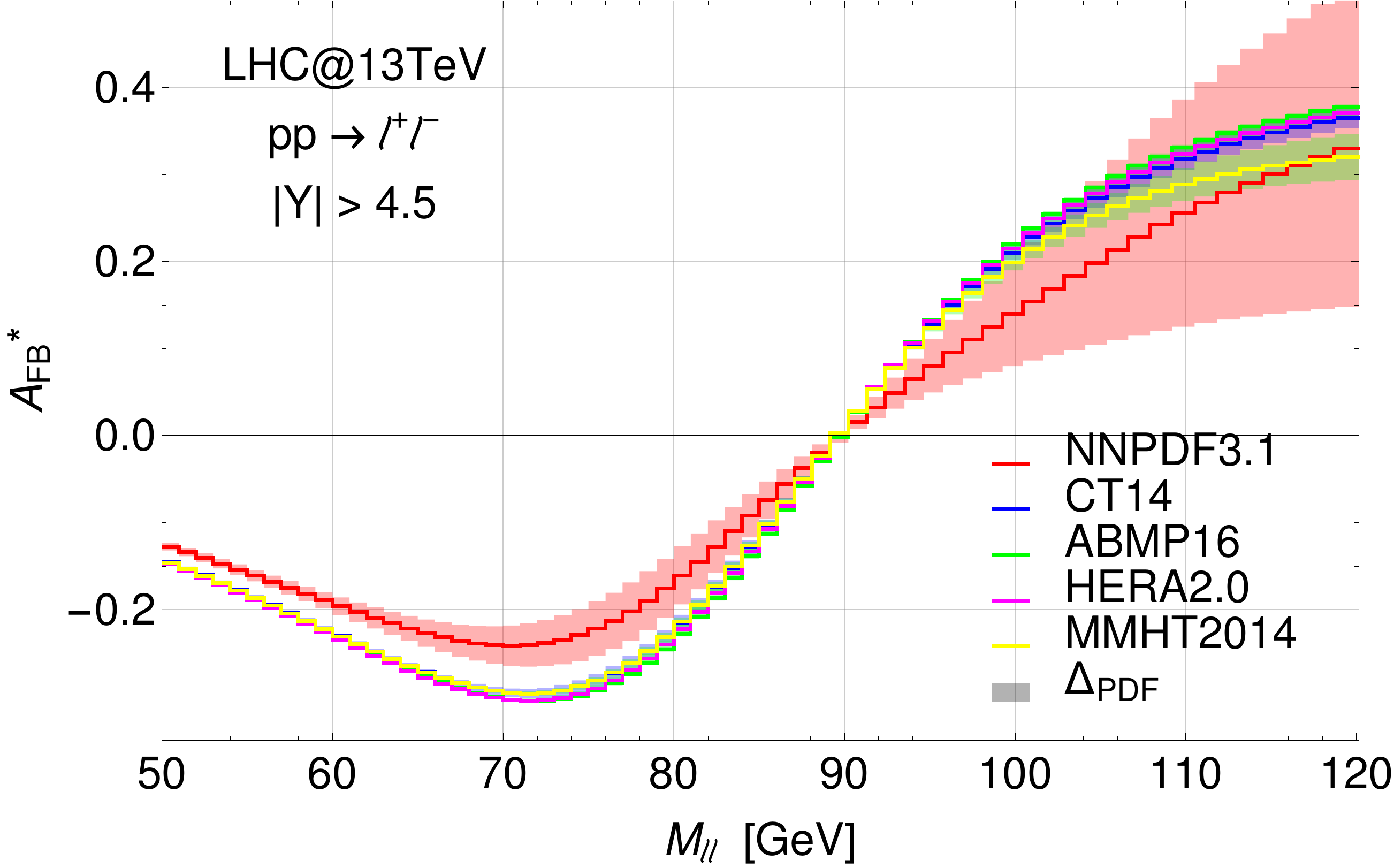}{(b)}
\caption{$A_{\rm FB}^*$ as a function of the di-lepton invariant mass including the (a) statistical and (b) PDF errors with a rapidity cut $|y|>4.5$. A luminosity $L = 300$ fb$^{-1}$ is assumed and acceptance cuts $|\eta| < 5$ and $p_T > 20$ GeV are imposed on both leptons.}
\label{fig:AFB_high_rapidity}
\end{center}
\end{figure}

In Fig.~\ref{fig:AFB_high_rapidity} we show the equivalent picture for the case of the $A_{\rm FB}^*$ observable.
In the left-hand side plot, we observe that the precision allowed by the available statistics is sufficient for a precise experimental measurement of the observable, reflecting what shown for the predicted di-lepton spectrum.
In the plot on the right-hand side, we can appreciate the good features of the $A_{\rm FB}^*$ in terms of reduction of the PDF uncertainties,
such that for this observable it will be possible to distinguish the predictions obtained with some of the PDF sets in specific invariant mass regions.
In particular, the NNPDF prediction cannot be disentangled from the others just by observing the differential cross section in the di-lepton invariant mass, because of the large PDF error which the spectrum is affected by.
In the $A_{\rm FB}^*$ distribution instead the NNPDF central value and error band come out to be incompatible with those of the other PDF sets.
A new sensitivity to the PDFs is then gained by associating the measurement of $A_{\rm FB}^*$ to the spectrum.

\section{Conclusions}
\label{sec:summa}
In summary, building upon the recent observation made in \cite{Accomando:2017scx} that the forward-backward asymmetry of di-lepton pairs produced via NC DY processes at the LHC can be used to constrain the PDFs inside the protons,
we emphasise here that its exploitation in conjunction with a cut on the rapidity of the final state gives access also to the flavour content of the proton.
In the presence of such a cut, one could indeed separate the valence/sea up-quark and the sea up-type antiquark, as the $y$-cut enforces a suppression of the $d\bar d$-quark luminosity.
Depending on the size of the data sample, the statistical and PDF errors for the different PDF sets could be not overlapping, thus offering the possibility to constrain or even eventually measure the up-type quark/antiquark PDFs. 

This should be intended as a preliminary analysis, quite promising though.
Higher order EW and QCD corrections can become quite important in the kinematical region of high rapidity.
They should be properly included in order to make solid statements.
Precise and reliable Monte Carlo Event Generators are available to this purpose and are regularly used by the LHC experimental collaborations for data interpretation.
Having these tools at hand, it would be highly advisable to perform a new PDFs fitting procedure aimed at establishing, in a firm way and by direct comparison,
whether the use of the reconstructed forward-backward asymmetry can give better results in terms of PDFs constraints with respect to the triple differential cross sections now in use.

\section*{Acknowledgements}
\noindent
This work is supported by the STFC grant number ST/P000711/1 and by BMBF under contract 05H15PMCCA. 
F.~H.~acknowledges the support and hospitality of the CERN Theory Division (Geneva), DESY (Hamburg) and the University of Basque Country (Bilbao), while part of this work was being done.
All authors acknowledge partial financial support through the NExT Institute. We thank J.~Huston and V.~Bertone for useful discussions. 

\bibliographystyle{apsrev4-1}
\bibliography{bib}

\end{document}